\documentclass[twocolumn,english,preprintnumbers,amsmath,amssymb,pre]{revtex4}

\usepackage{graphicx}
\usepackage{color}

\begin{document}

\title{Elastic properties and mechanical stability of bilayer graphene: \\
       Molecular dynamics simulations}
\author{Carlos P. Herrero}
\author{Rafael Ram\'irez}
\affiliation{Instituto de Ciencia de Materiales de Madrid,
         Consejo Superior de Investigaciones Cient\'ificas (CSIC),
         Campus de Cantoblanco, 28049 Madrid, Spain }
\date{\today}

\begin{abstract}
Graphene has become in last decades a paradigmatic example
of two-dimensional and so-called van-der-Waals layered materials,
showing large anisotropy in their physical properties. Here we study 
the elastic properties and mechanical stability of graphene bilayers
in a wide temperature range by molecular dynamics simulations.
We concentrate on in-plane elastic constants and compression modulus,
as well as on the atomic motion in the out-of-plane direction.
Special emphasis is placed upon the influence of anharmonicity
of the vibrational modes on the physical properties of bilayer 
graphene. We consider the excess area appearing in the presence of
ripples in graphene sheets at finite temperatures.
The in-plane compression modulus of bilayer graphene is found to
decrease for rising temperature, and results to be higher
than for monolayer graphene. We analyze the mechanical instability 
of the bilayer caused by an in-plane compressive stress. This defines 
a spinodal pressure for the metastability limit of the material,
which depends on the system size. Finite-size effects are described 
by power laws for the out-of-plane mean-square fluctuation, 
compression modulus, and 
spinodal pressure. Further insight into the significance of our 
results for bilayer graphene is gained from a comparison with data 
for monolayer graphene and graphite. 
\end{abstract}

\maketitle

\section{Introduction}

Over the last few decades there has been a surge of interest
in carbon-based materials with $sp^2$ orbital hybridization, 
such as fullerenes, carbon nanotubes, and
graphene \cite{ho00,ge07,ka07}, continuously enlarging
this research field beyond the long-known graphite.
In particular, bilayer graphene displays peculiar electronic 
properties, which have been discovered and thoroughly studied 
in recent years \cite{ce21,bh22}.
It presents unconventional superconductivity for stacking
of the sheets twisted relative to each other by a precise
small angle \cite{ca18,ya19}.
Such rotated graphene bilayers show magnetic properties that 
may be controlled by an applied bias voltage \cite{go17,sb18}.
Also, localized electrons are present in
the superlattice appearing in a moir\'e pattern, 
so that one may have a correlated insulator \cite{ca18b}.
Bilayer graphene displays ripples and out-of-plane
deformations akin to suspended monolayers \cite{me07b},
thus giving rise to a lack of planarity which may be
important for electron scattering \cite{gi10}.

A deep comprehension of thermodynamic properties of
two-dimensional (2D) systems has been a challenge in
statistical physics for many years \cite{sa94,ne04}.
This question has been mainly discussed in the field of
biological membranes and soft condensed matter \cite{ch15,ru12},
for which analyses based on models with realistic interatomic
interactions are hardly accessible.
In this context, graphene is a prototype crystalline membrane,
appropriate to study the thermodynamic stability of 2D materials.
This problem has been addressed in connection with anharmonic
effects, in particular with the coupling between
in-plane and out-of-plane vibrational modes \cite{am14,ra18b}.
Bilayer graphene is a well-defined two-sheet crystalline 
membrane, where an atomic-level
characterization is feasible, thereby permitting one
to gain insight into the physical properties of this type
of systems \cite{za10b,he19,ba11,am14,he20}.

Mechanical properties of graphene, including elastic constants,
have been studied by using several
theoretical \cite{mi08b,sa11,an12b,lo16}
and experimental \cite{le08,po15b,an15c,pa17,ni17,ga18} techniques.
These methods have been applied to analyze monolayer
as well as multilayer graphene, including the
bilayer \cite{li18b,an12b,ov18,ch21b,mo22}.
In this context,
a theory of the evolution of phonon spectra and elastic
constants from graphene to graphite was presented by
Michel and Verberck \cite{mi08b}.
In particular, for bilayer graphene on SiC,
Gao {\em et al.} \cite{ga18} have found a transverse stiffness
and hardness comparable to diamond.
More generally, mechanical properties of graphene and its
derivatives have been reviewed by Cao {\em et al.} \cite{ca18c},
and various effects of strain in this material were
reported by Amorim {\em et al.} \cite{am16}.

Several authors have addressed finite-temperature properties
of graphene using various kinds of atomistic
simulations \cite{fa07,ak12,ma14,lo16,ko16}.
In particular, this type of methods have been applied to study 
bilayer graphene \cite{za10b,li18b,he19,zh19,he20,ch21b}.
Thus, MD simulations were used to study mechanical 
properties \cite{zh19}, as well as the influence of
extended defects on the linear elastic constants of
this material \cite{li18b}.

In this paper we extend earlier work on isolated graphene sheets
to the bilayer, for 
which new aspects show up due to interlayer interactions
and the concomitant coupling between atomic displacements
in the out-of-plane direction.
We use molecular dynamics (MD) simulations to study structural 
and elastic properties of bilayer graphene at temperatures up 
to 1200~K. Especial emphasis is laid on the behavior of
bilayer graphene under tensile in-plane stress and on its
mechanical stability under compressive stress.
MD simulations allow us to approach the spinodal line in
the phase diagram of bilayer graphene, which defines its
stability limit.
We compare results found for the bilayer with data corresponding
to monolayer graphene and graphite, which yields information
on the evolution of physical properties from an individual
sheet to the bulk.

The paper is organized as follows. In Sec.~II we describe the
method employed in the MD simulations.
In Sec.~III we present the phonon dispersion bands and the
elastic constants at $T = 0$.
In Sec.~IV we present results for structural properties 
derived from the simulations:
interatomic distances, interlayer spacing, and 
out-of-plane atomic displacements.
The in-plane and excess area are discussed in Sec.~V,
and in Sec.~VI we analyze the elastic constants and
compressibility at finite temperatures, along with
the stability limit for compressive stress.
Finite-size effects are studied in Sec.~VII.
The papers closes with a summary of the main results in
Sec.~VIII.

\section{Method of calculation}

In this paper we employ MD simulations to study structural and 
elastic properties of graphene bilayers as functions of temperature
and in-plane stress.
The interatomic interactions in graphene are described with a 
long-range carbon bond-order potential, the so-called 
LCBOPII \cite{lo05}, used earlier to perform simulations 
of carbon-based systems, such as graphite \cite{lo05}, 
diamond,\cite{lo05} and liquid carbon \cite{gh05}.
In more recent years, this interatomic potential has been utilized 
to study graphene \cite{fa07,ra16,za10b,lo16}, and in particular 
mechanical properties of this 2D material \cite{za09,ra17}.
The LCBOPII potential model was also used to conduct
quantum path-integral MD simulations of graphene monolayers \cite{he16} 
and bilayers \cite{he19}, which allowed to assess nuclear quantum 
effects in various properties of this material.
Here, as in earlier simulations \cite{ra16,he16,ra17},
the original LCBOPII parameterization has been slightly modified
in order to rise the bending constant $\kappa$ of a graphene monolayer 
from 0.82 eV to a value of 1.49 eV, close to experimental results and
{\em ab-initio} calculations \cite{la14}. Values of the parameters 
employed here for the torsion term of the potential are given
in Appendix A.1.

For the interlayer interaction we have considered the same parameterization
as that previously used in simulations of graphene bilayers with this 
potential model \cite{za10b,he19}, presented in Appendix A.2.
For the minimum-energy configuration of bilayer graphene with AB 
stacking, we find an interlayer binding energy of 25 meV/atom
(50 meV/atom for graphite) \cite{za10b}.

Our simulations were carried out in the 
isothermal-isobaric ensemble, where one fixes the number of carbon 
atoms, $2N$ (i.e., $N$ atoms per sheet), the in-plane stress tensor,
$\{ \tau_{ij} \}$, and the temperature, $T$.
We have considered rectangular supercells with similar side lengths
in the $x$ and $y$ directions in the layer plane, $L_x \approx L_y $.
These supercells included from $N$ = 48 to 8400 carbon atoms per 
graphene sheet.
Periodic boundary conditions were assumed for $x$ and $y$ coordinates, 
whereas C atoms were allowed to move freely in the out-of-plane 
$z$ coordinate (free boundary conditions). 

To keep a given temperature $T$, chains of four Nos\'e-Hoover thermostats
were connected to each atomic degree of freedom \cite{tu98}.
An additional chain including four thermostats was connected to the 
barostat which regulates the in-plane area of the simulation cell 
($xy$ plane), keeping the required stress 
$\{ \tau_{ij} \}$ \cite{tu98,al87}.
Integration of the equations of motion was performed by using
the reversible reference system propagator algorithm (RESPA), which
permits to consider different time steps for slow and fast degrees
of freedom \cite{ma96}.
For the atomic dynamics derived from the LCBOPII potential,
we took a time step $\Delta t$ = 1 fs, which gave good accuracy
for the temperatures and stresses discussed here.
For fast dynamical variables such as the thermostats, we used
$\delta t = \Delta t/4$.

\begin{figure}
\vspace{-0.0cm}
\includegraphics[width= 7cm]{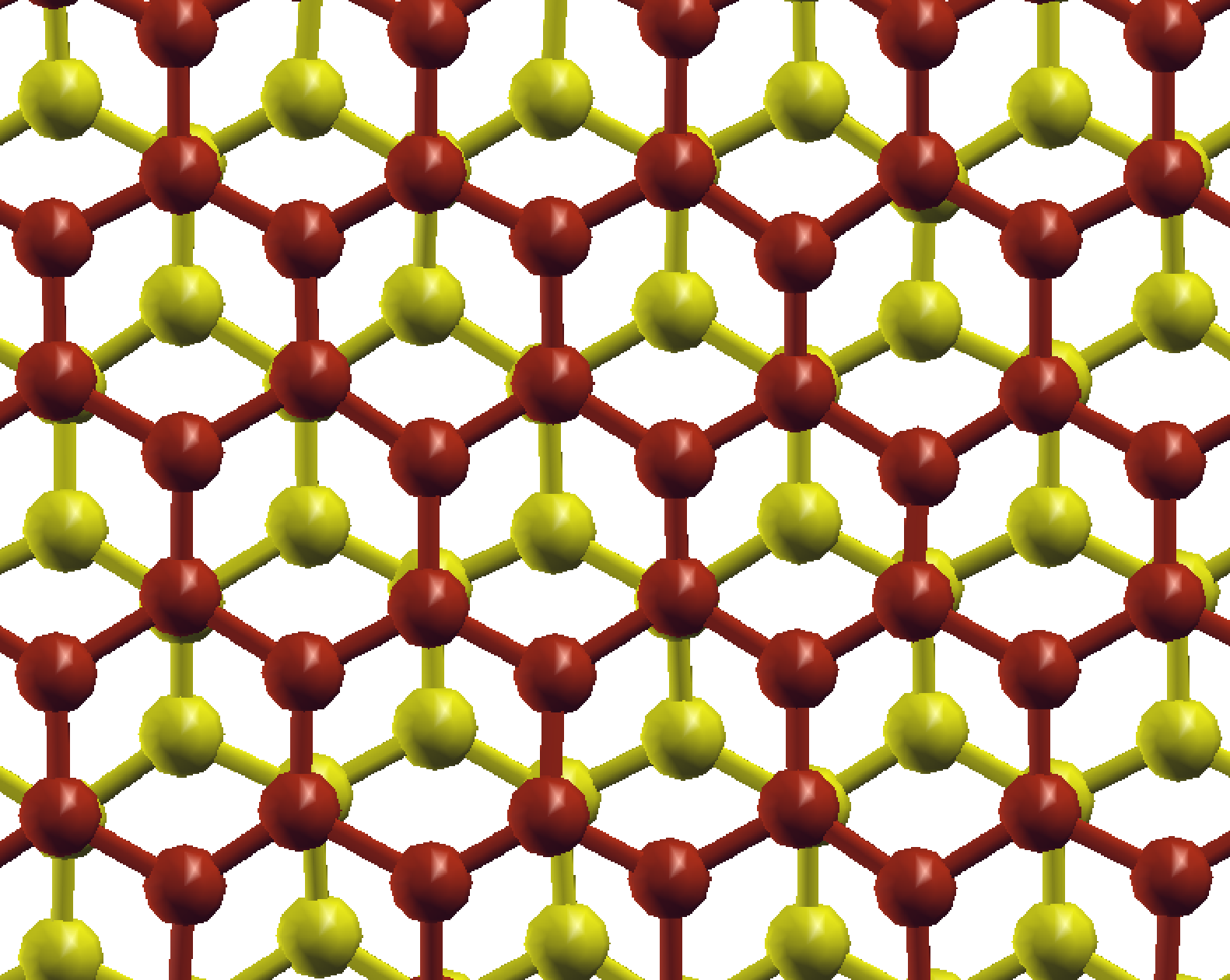}
\includegraphics[width= 7cm]{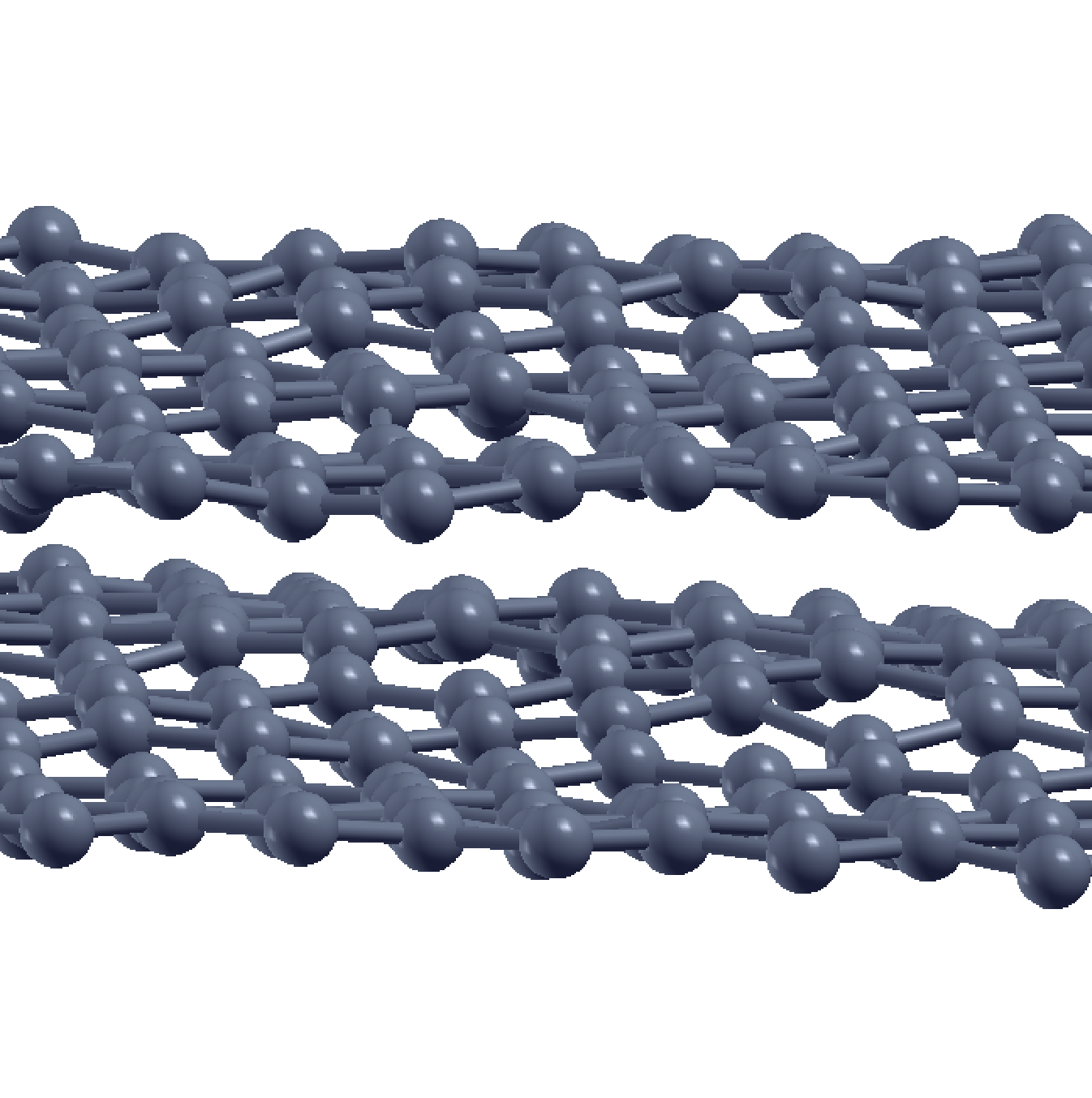}
\caption{Top (upper) and side (lower) views of an atomic configuration
of bilayer graphene obtained from MD simulations at $T$ = 800 K.
In the top view, red and yellow spheres represent carbon atoms in
the upper and lower graphene sheets, respectively.
}
\label{f1}
\end{figure}

The configuration space has been sampled for $T$ in the range
from 50 to 1200 K.
Given a temperature, a typical run consisted of $2 \times 10^5$ MD
steps for system equilibration and $8 \times 10^6$ steps
for calculation of ensemble averages.
In Fig.~1 we show top and side views of an atomic configuration 
of bilayer graphene obtained in MD simulations at $T = 800$~K.
In the top view, red and yellow balls stand for carbon atoms in 
the upper and lower graphene layers in AB stacking pattern.
In the side view, one can see atomic displacements in the
out-of-plane direction, clearly observable at this temperature.

To characterize the elastic properties of bilayer graphene 
we consider uniaxial stress along the $x$ or $y$ directions, 
i.e., $\tau_{xx} \neq 0$ or $\tau_{yy} \neq 0$, as well as
2D hydrostatic pressure $P$ (biaxial stress) \cite{be96b},
which corresponds to 
$\tau_{xx} =  \tau_{yy} = -P$, $\tau_{xy} = 0$. 
Note that $P > 0$ and $P < 0$
mean compressive and tensile stress, respectively.

For comparison with our results for graphene bilayers, we also
performed some MD simulations of graphite using the same 
potential LCBOPII. In this case we considered cells containing
$4 N$ carbon atoms (four graphene sheets), and periodic
boundary conditions were assumed in the three space directions.

Other interatomic potentials have been used in last years 
to analyze several properties of graphene, in particular the
so-called AIREBO potential \cite{sf15,me15,zo16,an16,gh17}.
Both LCBOPII and AIREBO models give very similar values for 
the equilibrium C--C interatomic distance and for the thermal 
expansion coefficient \cite{me15,gh17,he16}.
For the Young's modulus of graphene, we find that the result obtained
by employing the LCBOPII potential is closer to those yielded by 
{\em ab initio} calculations \cite{me15}.

\section{Phonon dispersion bands and elastic constants at $T = 0$}

The elastic stiffness constants, $c_{ij}$, of bilayer graphene calculated
with the LCBOPII potential model in the limit $T \rightarrow 0$ 
can be used as reference values for the finite-temperature analysis
presented below. These elastic constants are calculated here from
the harmonic dispersion relation of acoustic phonons.
The interatomic force constants utilized to obtain the dynamical
matrix were obtained by numerical differentiation
of the forces using atom displacements of $1.5\times10^{-3}$~\AA\
with respect to the equilibrium (minimum-energy) positions.

\begin{figure}
\includegraphics[width=7cm]{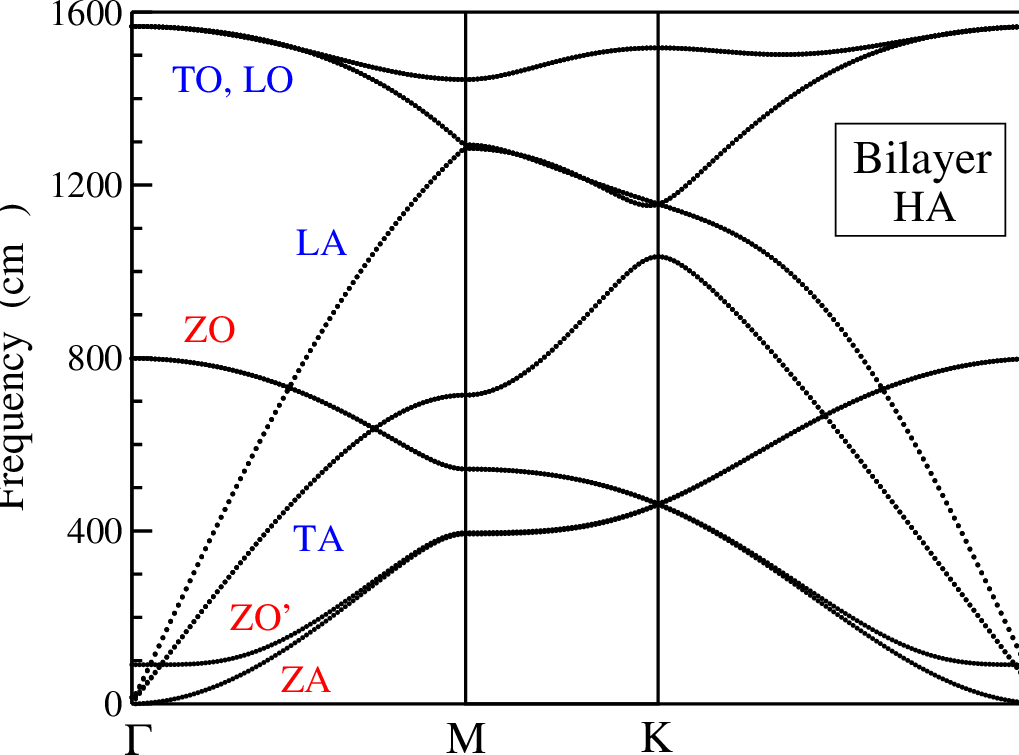}
\vspace{0.3cm}
\caption{Phonon dispersion bands of bilayer graphene, obtained from
diagonalization of the dynamical matrix for the LCBOPII
potential model.
}
\label{f2}
\end{figure}

The phonon dispersion of bilayer graphene, calculated by 
diagonalization of the dynamical matrix is presented in Fig.~2 
along high-symmetry directions of the 2D Brillouin zone.
One finds 12 phonon bands, corresponding to four C atoms 
(2 per layer) in the crystallographic unit cell. Labels indicate 
the common names of the phonon bands:
eight branches with in-plane atomic motion (LA, TA, LO, and TO,
all of them two-fold degenerate),
and four branches with displacements along the out-of-plane direction 
(ZA, ZO', and a two-fold degenerate ZO band).
The phonon dispersion presented in Fig.~2 is analogous to those 
obtained for other empirical potentials and DFT 
calculations \cite{ya08,ka11,ko15b,si13b}.
We emphasize the presence of the flexural 
ZA band, which is parabolic close to the $\Gamma$ point 
($\omega \sim k^2$), and typically appears in 
2D materials \cite{wi04,am13,ra19,he21b}.
Here $k$ denotes the wavenumber, i.e., $k = |{\bf k}|$, and 
${\bf k} = (k_x, k_y)$ is a wavevector in the 2D hexagonal Brillouin zone.

Note also the presence of the optical mode ZO', which does not appear
in monolayer graphene, and in the case of the bilayer corresponds to 
the layer-breathing Raman-active $A_{2g}$ mode, for which
a frequency of 89 cm$^{-1}$ has been measured \cite{li18}.
The LCBOPII potential yields for this band at the $\Gamma$ point
($k = 0$) a frequency of 92 cm$^{-1}$.
This value is close to that found from {\em ab initio} calculations for 
graphene bilayers \cite{ya08}.

The interatomic potential LCBOPII was used before to calculate
the phonon dispersion bands of graphene and graphite \cite{ka11}.
However, the version of the potential employed in 
Ref.~\cite{ka11} was somewhat different from that considered 
here, which gives a description of the graphene bending closer to 
experimental results \cite{la14,ra16} (see Appendix A.1).

The sound velocities for the acoustic bands LA and TA
along the direction $\Gamma\mathrm{M}$,
with wavevectors $(k_x,0,0)$, are given by the slope
$\left( \partial \omega / \partial k_x \right)_{\Gamma}$
in the limit $k_x \to 0$.
The elastic stiffness constants can be obtained from these
velocities by using the following expressions, valid for the
hexagonal symmetry of graphene \cite{ne05}:
\begin{equation}
  c_{11} = \rho \left( \frac{\partial \omega_{\rm LA}}{\partial k_x }  
	   \right)^2_{\Gamma}  \; ,
\label{c11}
\end{equation}
\begin{equation}
  c_{12} = c_{11} - 2 \rho \left( \frac{\partial \omega_{\rm TA}}
	   {\partial k_x }  \right)^2_{\Gamma}  \; .
\label{c12}
\end{equation}
where $\rho$ is the surface mass density of graphene.
We find $c_{11}$ = 20.94 eV \AA$^{-2}$ and 
$c_{12}$ = 4.54 eV \AA$^{-2}$. Note that the dimensions of these 
elastic constants (force/length) coincide with those of 
the in-plane stress.
These $c_{ij}$ can be converted into elastic constants $C_{ij}$
(units of force per square length), typical of 
three-dimensional (3D) materials,
as $C_{ij} = c_{ij} / d_0$, using the interlayer distance $d_0$ of
the minimum-energy configuration of bilayer graphene.
Taking $d_0 = 3.3372$ \AA, we find for the bilayer $C_{11}$ = 1005~GPa 
and $C_{12}$ = 218~GPa, near the values found for graphite in the
classical low-$T$ limit, using the LCBOPII potential: 1007 and 
216~GPa, respectively \cite{he21b}.

\section{Structural properties}

\subsection{Interatomic distance}

For bilayer graphene, the minimum-energy configuration for the
LCBOPII potential corresponds
to planar sheets and the interatomic distance between nearest neighbors 
in a layer amounts to 1.4193~\AA. This distance turns out to be 
a little smaller than that
found for monolayer graphene using the same interatomic
potential ($r_0$ = 1.4199 \AA). This fact was noticed by
Zakharchenko {\em et al.} \cite{za10b} in their results of Monte
Carlo simulations of graphene bilayers.

\begin{figure}
\includegraphics[width=7cm]{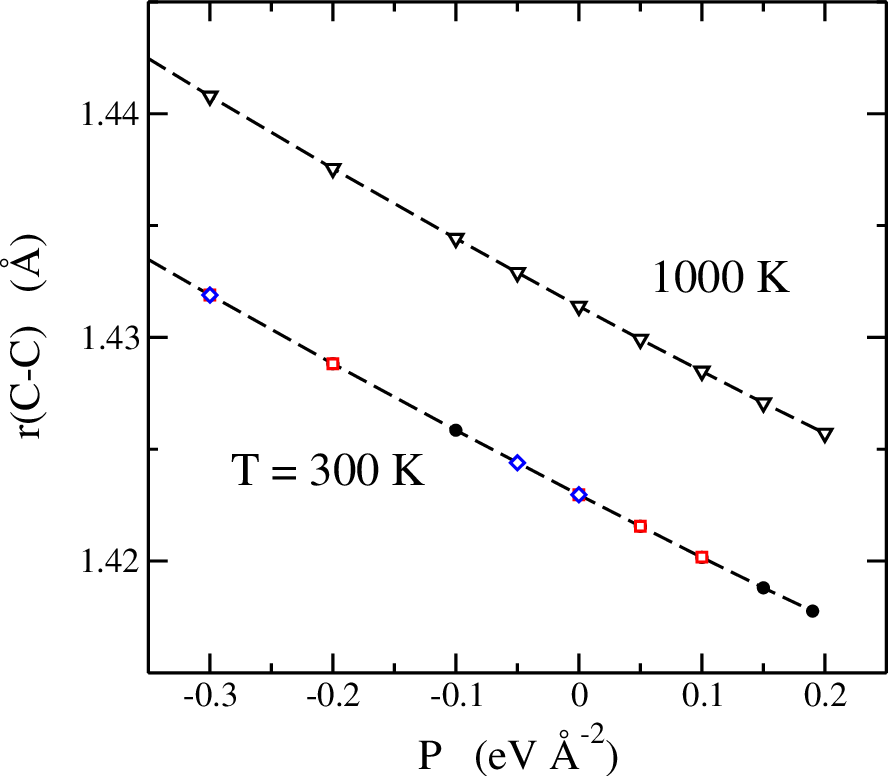}
\vspace{0.3cm}
\caption{Interatomic distance vs 2D hydrostatic pressure for several
cell sizes at $T$ = 300 and 1000 K.
Data points are derived from MD simulations.
For $T$ = 300 K, the data correspond to cell sizes:
$N$ = 240 (solid circles), 448 (open squares), and
3840 (open diamonds). Open triangles correspond to $N$ = 240 at
1000 K. Error bars are less than the symbol size.
}
\label{f3}
\end{figure}

We have studied the change of the interatomic C--C distance $r$
(actual distance in 3D space)
as a function of 2D hydrostatic pressure, $P$,  at several temperatures.
In Fig.~3 we present the dependence of $r$ on $P$ at $T$ = 300 and 1000~K. 
For $T$ = 300 K, data derived from MD simulations are shown for three
cell sizes : $N$ = 240 (solid circles), 448 (open squares), and
3840 (open diamonds). We observe that the size effect on the interatomic 
distance is negligible, since differences between the results for 
different cell sizes are much smaller than the symbol size in Fig.~3.
The data for $T$ = 1000~K (open triangles) correspond to $N$ = 240.
Note that 2D hydrostatic pressure $P > 0$ corresponds to compressive
stress.

Close to $P = 0$,
this dependence can be fitted for $T$ = 300 K to an expression
$r = r_m + \mu P$, where $r_m$ = 1.4230 \AA\ is the interatomic 
distance for the stress-free bilayer at this temperature and 
$\mu = -0.0289$ \AA$^3$/eV. 
For $T$ = 1000~K, we find $r_m$ = 1.4314 \AA\ and
$\mu = -0.0302$~\AA$^3$/eV. This slope is slightly larger
than that found for $T$ = 300 K.

In connection with the interatomic distance $r$, we note that for
a strictly planar geometry, the area per atom for an ideal 
honeycomb pattern is given by $S_p = 3 \sqrt{3} \, r^2 / 4$.
At finite temperatures, however, the graphene layers are not 
totally planar and the actual in-plane area per atom is smaller 
than that given by the above expression, using the mean 
interatomic distance between nearest-neighbor C atoms.
This is related with the so-called {\em excess area} and is
discussed below in Sec.~V.

In each hexagonal ring, two C--C bonds are aligned parallel to
the $y$ direction (vertical in Fig.~1, top image), and the four other
bonds form an angle of 30 degrees with the $x$ axis
(horizontal direction). A compressive uniaxial stress along 
the $y$ axis ($\tau_{yy} < 0$) causes a decrease in the length of the
former bonds and an increase in the latter, as corresponds
to a positive Poisson's ratio. The opposite happens 
for $\tau_{xx} < 0$.

\subsection{Interlayer spacing}

For the minimum-energy configuration we find an interlayer distance
$d_0 = 3.3372$ \AA, to be compared with a distance of 3.3371~\AA\
obtained in Ref.~\cite{za10b} using an earlier version of the
LCBOPII potential. At $T$ = 300~K we obtain $d$ = 3.374~\AA,
i.e., the interlayer distance increases somewhat due to 
bending of the graphene sheets caused by thermal motion.

The interlayer spacing is reduced in the presence of a tensile 
2D hydrostatic pressure. Thus, for $P = -0.5$ eV/\AA$^2$ and 
$T$ = 300 K, we find $d$ = 3.367~\AA. 
This decrease is due to a reduction in the out-of-plane
fluctuations under a tensile stress. 
The effect of this relatively high stress on the distance $d$ is,
however, smaller than the thermal expansion up to 300~K.
In the presence of a compressive stress in the $xy$ plane, 
one has an expansion of the interlayer spacing, but this kind of 
stress causes an instability of the bilayer configuration for 
relatively small values of $P$, as explained below.

\begin{figure}
\includegraphics[width=7cm]{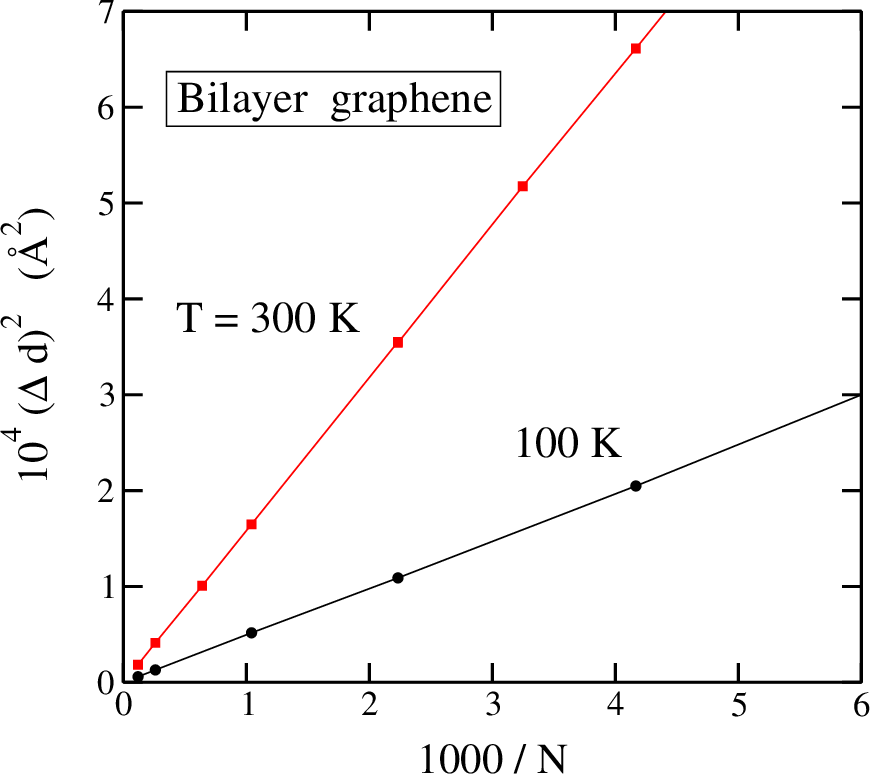}
\vspace{0.3cm}
\caption{MSF of the interlayer distance as a function of the inverse
cell size for $T$ = 100 K (circles) and 300 K (squares), as derived
from MD simulations of bilayer graphene.
}
\label{f4}
\end{figure}

The mean-square fluctuation (MSF) of the interlayer distance, 
$(\Delta d)^2 = \langle d^2 \rangle - \langle d \rangle^2$, 
associated to thermal motion at finite temperatures, is related 
to the interaction between graphene layers. This MSF is
expected to depend on the size of the considered simulation
cell, and becomes negligible in the thermodynamic limit ($N \to \infty$).
In Fig.~4 we display $(\Delta d)^2$ derived from our MD simulations 
as a function of the inverse cell size for stress-free bilayer 
graphene at $T = 100$~K (circles) and 300~K (squares). 
One observes that $(\Delta d)^2 \to 0$ for $1/N \to 0$, and
grows linearly for increasing inverse cell size.

To connect these results with the energetics of bilayer graphene,
we have calculated the interlayer interaction energy 
for several values of the spacing $d$ near the distance $d_0$ corresponding 
to the minimum-energy configuration.  The interaction energy per atom 
can be written as:
\begin{equation}
    E_{\rm int} = E_{\rm int}^0 + \frac12 k (d - d_0)^2  \, .
\label{ed}
\end{equation}
where $E_{\rm int}^0$ is the energy for distance $d_0$, and $k$ is
an effective interaction constant which is found to amount to
0.093 eV/\AA$^2$.
Then, for the whole simulation cell ($2N$ atoms in a bilayer), 
the energy corresponding to a distance $d$ close to $d_0$ is
\begin{equation}
  \bar{E}_{\rm int} = \bar{E}_{\rm int}^0 + k N (d - d_0)^2  \, .
\label{eint}
\end{equation}
Thus, thermal motion at temperature $T$, associated to
the degree of freedom $d$, will cause a MSF of this variable,
$(\Delta d)^2$, given by the mean potential energy:
\begin{equation}
   k N (\Delta d)^2  =  \frac12 k_B T   \, ,
\label{knd}
\end{equation}
where $k_B$ is Boltzmann's constant. 
This means that for a given temperature, $(\Delta d)^2$ scales
as $1/N$, as shown in Fig.~4.

As indicated in Sec.~III, the phonon spectra of monolayer and 
bilayer graphene are similar. The main difference between them is 
the appearance in the latter of the 
ZO' vibrational band, which is almost flat in a region of
${\bf k}$-space near the $\Gamma$ point (see Fig.~2).
As noted above, this vibrational mode of bilayer graphene is
the layer-breathing Raman-active $A_{2g}$ mode \cite{li18}.
The frequency of the ZO' band at $\Gamma$ (which will be denoted 
here $\omega_0$) can be related to the interlayer coupling constant 
$k$ as  $\omega_0 = (k_N / M_{\rm red})^{1/2}$,
with $k_N = 2 N k$ and the reduced mass $M_{\rm red} = N m / 2$
($m$: atomic mass of carbon). We find $\omega_0 = 2 (k / m)^{1/2}$. 
and putting for the coupling constant $k$ = 0.093 eV \AA$^{-2}$/atom, 
one obtains $\omega_0 =$ 92 cm$^{-1}$, which coincides
with the frequency of the ZO' band derived from the dynamical matrix
at the $\Gamma$ point.
Michel and Verberck \cite{mi08b} have studied the evolution of
this frequency $\omega_0$ with the number of sheets $n$ in graphene
multilayers. They found an increase of $\omega_0$ for rising
$n$, which saturates to a value of 127 cm$^{-1}$ for 
large $n$ (graphite).

The interlayer coupling in bilayer graphene was studied before by 
Zakharchenko {\em et al.} \cite{za10b} by means of Monte Carlo 
simulations.  The low-frequency part of the ZO' band was described 
by these authors using a parameter $\gamma$, which is related to 
the parameters employed here as $\gamma = \rho \, \omega_0^2 / 4$,
where $\rho$ is the surface mass density. 
From this expression we find $\gamma$ = 0.035 eV \AA$^{-4}$,
which agrees with the low-temperature result derived from 
Monte Carlo simulations (Fig.~7 in Ref.~\cite{za10b}).

Fluctuations in the interlayer spacing of bilayer graphene at
temperature $T$ are related with the isothermal compressibility 
in the out-of-plane direction, $\chi_z$ \cite{he19}.
In fact, $\chi_z$ can be calculated from the MSF $(\Delta d)^2$
by using the expression \cite{he19}
\begin{equation}
   \chi_z = \frac{L_x L_y}{k_B T} 
      \frac{(\Delta d)^2}{\langle d \rangle}   \, .
\label{chiz}
\end{equation}
Using $(\Delta d)^2 = 1.65 \times 10^{-4}$~\AA$^2$ for $N$ = 960 at
$T$ = 300~K, we find $\chi_z = 2.96 \times 10^{-2}$ GPa$^{-1}$.
This value is a little larger than experimental results for
graphite, of about $2.7 \times 10^{-2}$ GPa$^{-1}$ \cite{bl70,ni72}.
This is consistent with the fact that bilayer graphene is more 
compressible than graphite in the $z$ direction, since in the
latter each layer is surrounded by two other graphene layers,
whereas in the former each sheet has a single neighbor.

\subsection{Out-of-plane motion}

The minimum-energy configuration for the graphene layers,
i.e., the classical low-temperature limit, corresponds to
planar sheets. At finite temperatures the graphene sheets are 
bent.  This bending is directly related to the atomic motion 
in the out-of-plane $z$ direction, whose largest vibrational
amplitudes come from low-frequency ZA modes with long wavelength 
(small $k$).
For stress-free graphene, the ZA phonon branch can be described
close to the $\Gamma$ point by a parabolic dispersion relation of 
the form $\omega({\bf k}) = \sqrt{\kappa/\rho} \, k^2$,
with the bending constant $\kappa$ = 1.49 eV (see Fig.~2).

It is known that the atomic MSF in the $xy$ layer plane is
relatively insensitive to the system size, but the out-of-plane 
MSF has important finite size effects.
This dependence on $N$ has been studied earlier for
stress-free monolayer and bilayer graphene by means of Monte Carlo 
and molecular dynamics simulations, in particular using the LCBOPII 
interatomic potential \cite{he16}.
For system size $N$, one has an effective cut-off for
the wavelength $\lambda$ given by $\lambda_{\rm max} \approx L$,
where $L = (N S_p)^{1/2}$, and $S_p$ is the in-plane area per atom.
Thus, the minimum wavenumber present for size $N$ is
$k_{\rm min} = 2 \pi / \lambda_{\rm max}$, and the minimum frequency
for ZA modes is
\begin{equation}
   \omega_{\rm min} \approx  \frac{4 \pi^2}{N S_p}  \, 
	      \left( \frac{\kappa}{\rho}  \right)^{\frac12}  \; ,
\end{equation}
so that $\omega_{\rm min}$ scales as $N^{-1}$.

\begin{figure}
\includegraphics[width=7cm]{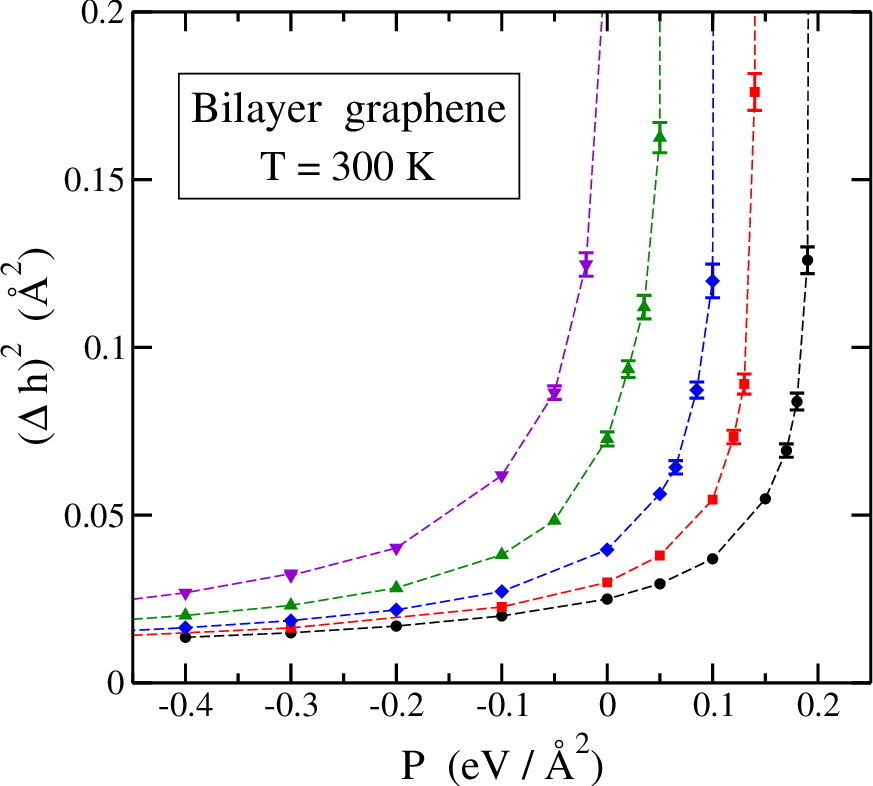}
\vspace{0.3cm}
\caption{Atomic MSF in the $z$ direction vs 2D hydrostatic pressure
for several cell sizes at $T$ = 300 K.
From left to right: $N$ = 3840 (triangles down), 960 (triangles up),
448 (diamonds), 308 (squares), and 240 (circles).
Positive (negative) $P$ means compressive (tensile) stress.
Dashed lines are guides to the eye.
}
\label{f5}
\end{figure}

For a graphene sheet, we call ${\bf r} \equiv (x, y)$ the 2D position 
on the layer plane and $h({\bf r})$ is the distance to the mean plane 
of the sheet.
In Fig.~5 we present the MSF of the atomic positions in the $z$ direction
for bilayer graphene,
$(\Delta h)^2 = \langle h^2 \rangle - \langle h \rangle^2$,
as a function of 2D hydrostatic pressure $P$ for various cell sizes at 
$T$ = 300 K. Symbols represent results derived from our MD simulations,
with cell size decreasing from left to right: $N$ = 3840, 960, 448,
308, and 240. 
One observes first that $(\Delta h)^2$ appreciably increases for
rising system size, as expected from earlier studies of 
2D materials \cite{ga14}.
For the largest size displayed in Fig.~5, $N = 3840$, we find at
$P = 0$, $(\Delta h)^2$ = $0.22 \pm 0.01$ \AA$^2$
(not shown in the figure).
The dependence of $(\Delta h)^2$ on $N$ for stress-free bilayer 
graphene will be analyzed below in Sec.~VII.
Second, we also observe in Fig.~5 that the difference in atomic MSF 
between different system sizes is reduced for increasing tensile 
stress ($P < 0$).  $(\Delta h)^2$ grows as 
the tensile stress is reduced ($d (\Delta h)^2 / d P > 0$), and 
eventually diverges at a size-dependent critical pressure $P_c(N) > 0$.
Third, one sees that $P_c(N)$ approaches zero for rising system size.

The dependence of the critical pressure $P_c$ on $N$ will be 
discussed below in relation to fluctuations of the in-plane area 
$S_p$, which are also found to diverge in parallel with $(\Delta h)^2$ 
for each size $N$.
The origin of this instability is related to the appearance
of imaginary frequencies for vibrational modes in the ZA flexural
band for pressure $P_c(N)$.
This will be discussed in Sec.~VI in connection with the 
2D modulus of hydrostatic compression $B_p$, which is found
to vanish at $P_c$.

\section{In-plane and excess area}

The in-plane area, $S_p = L_x L_y / N$, is the variable conjugate to 
the pressure $P$ in the isothermal-isobaric ensemble considered here. 
Its temperature dependence, $S_p(T)$, has been analyzed earlier in 
detail for monolayer and bilayer graphene from atomistic 
simulations \cite{za10b,he16,he19}.
For the bilayer we find in the minimum-energy configuration (T = 0)
an area $S_0$ = 2.6169 \AA$^2$/atom, in agreement with earlier
calculations \cite{za10b,he19}.
Here, we discuss the behavior of $S_p$ as a function
of 2D hydrostatic stress, both tensile and compressive.

\begin{figure}
\includegraphics[width=7cm]{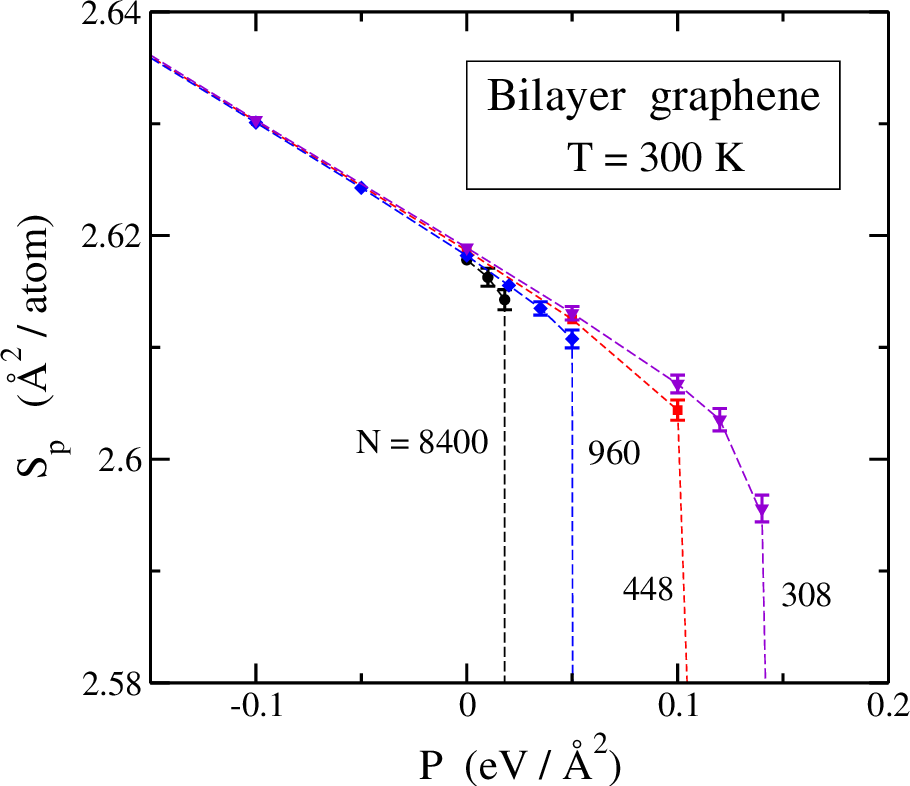}
\vspace{0.3cm}
\caption{In-plane area per atom vs 2D hydrostatic pressure $P$ at
$T$ = 300~K. Shown are results for cell sizes $N$ = 308 (traingles),
448 (squares), 960 (diamonds), and 8400 (circles).
Error bars, when not displayed, are in the order of the symbol size.
Dashed lines are guides to the eye.
}
\label{f6}
\end{figure}

In Fig.~6 we display the dependence of $S_p$ on $P$ for several
cell sizes at $T$ = 300~K. We present data for $N$ = 308, 448, 960,
and 8400. For tensile stress $P < -0.05$~eV/\AA$^2$, 
$S_p$ data for different
cell sizes are indistinguishable at the scale of Fig.~6. In fact, 
we obtain a nearly linear dependence with a slope 
$d S_p / d P \approx -0.12$~\AA$^4$/eV. However, differences
appear close to $P = 0$, and even more for
compressive stress ($P > 0$).  For each size $N$, one observes
a fast decrease in $S_p$ close to the corresponding stability
limit of the planar phase. We obtain values of the in-plane area
below 2.58 \AA$^2$/atom, not shown in the figure for clarity.
Changes in $S_p$ correspond to linear strain $\epsilon_L$ as:
$S_p = S_0 (1 + \epsilon_L)^2$. This means that the vertical range
in Fig.~6 corresponds to a strain range between 
$\epsilon_L = -7.1 \times 10^{-3}$ (compression) and 
$4.4 \times 10^{-3}$ (tension).

In our MD simulations, carbon atoms are free to move in the out-of-plane
direction ($z$ coordinate), and the {\em real} surface
of a graphene sheet is not strictly planar, having an actual area
larger than the area of the simulation cell in the $xy$ plane.
Differences between the {\em in-plane} area $S_p$ and {\em real} area 
$S_r$ were considered earlier in the context of biological
membranes \cite{im06,wa09,ch15} and in recent years for graphene, as
a paradigmatic crystalline membrane \cite{ra17,ni17}.
An explicit differentiation between both areas is relevant to understand 
certain properties of 2D materials \cite{sa94}.
Some experimental techniques can be sensitive to properties connected
to the area $S_r$, whereas other methods may be suitable to analyze
variables related to the area $S_p$ \cite{ni15,ni17}.

Here we calculate the real area $S_r$ of both sheets in bilayer graphene
by a triangulation method based on the atomic positions along
the simulation runs \cite{ra17,he19}.
In the sequel, $S_r$ will denote the real area per atom.
The difference $S_r - S_p$ has been called in the literature 
{\em hidden} area \cite{ni17} or {\em excess} area \cite{he84,fo08}.
We consider the dimensionless {\em excess} area $\Phi$
for a graphene sheet, defined as \cite{he84,fo08}
\begin{equation}
    \Phi = \frac{S_r - S_p}{S_p}  \; .
\end{equation}
In the classical low-temperature limit, $\Phi$ vanishes,
as the sheets become strictly planar for $T \to 0$. We note
that this is not the case in a quantum calculation, where one has 
$\Phi > 0$ for $T \to 0$ due to atomic zero-point motion \cite{he18}.

The excess area is related to the amplitude of the vibrational 
modes in the $z$ direction. This allows one to find analytical 
expressions for $\Phi$ in terms of the frequency of those
modes.  The instantaneous real area $S_r^*$ may be expressed in
a continuum approximation as \cite{im06,wa09,ra17}
\begin{equation}
   S_r^* = \frac1N \int_0^{L_x} \int_0^{L_y} dx \, dy \, 
             \sqrt{1+|\nabla h({\bf r})|^2}  \; ,
\label{ains}
\end{equation}
where $h({\bf r})$ represents the distance to the mean $xy$ plane 
of the sheet, as in Sec.~IV.C.
Expanding $h({\bf r})$ as a Fourier series with wavevectors
${\bf k} = (k_x, k_y)$ in the 2D hexagonal Brillouin zone, the real 
area $S_r = \langle S_r^* \rangle$ may written as \cite{sa94,ch15,ra17}
\begin{equation}
   S_r = S_p  \left[ 1 + \frac{1}{2N}  \sum_{\bf k}  k^2
             \langle |H({\bf k})|^2 \rangle  \right]   \; ,
\label{aains}
\end{equation}
where $H({\bf k})$ are the Fourier components of $h({\bf r})$ 
(see Appendix B).
Taking into account that the MSF of a mode with frequency 
$\omega_j({\bf k})$ is given by $k_B T / m \omega_j({\bf k})^2$,
$m$ being the atomic mass of carbon, one finds for the excess area
\begin{equation}
 \Phi  =   \frac{k_B T}{2 m N}
        \sum_{j,{\bf k}} \frac{k^2}{\omega_j({\bf k})^2} \; ,
\label{omega3}
\end{equation}
where the sum in $j$ is extended to phonon branches with atomic
motion in the $z$ direction, i.e., ZA, ZO, and ZO'. 

For small $k$, the contribution of ZO and ZO' modes to the sum in
Eq.~(\ref{omega3}) vanishes for $k \to 0$, as in both cases
$\omega_j({\bf k})$ converges to positive values.
For the flexural ZA band with negligible effective stress,
we have  $\omega_{\rm ZA}({\bf k}) \propto k^2$,
and  $k^2 / \omega_{\rm ZA}({\bf k})^2  \propto  k^{-2}$,
so that the contribution of ZA modes with small $k$ is dominant
in the sum in Eq.~(\ref{omega3}).
The minimum wavenumber $k_{\rm min}$ available for cell size $N$ 
scales as $k_{\rm min} \sim N^{-1/2}$ (see Sec.~IV.C). Thus, its 
contribution to $\Phi$ grows linearly with $N$, and diverges 
for stress-free graphene in the thermodynamic limit. 
This divergence disappears in the presence of a tensile in-plane 
stress (even small), be it caused by internal tension or by an 
external pressure.

\begin{figure}
\includegraphics[width=7cm]{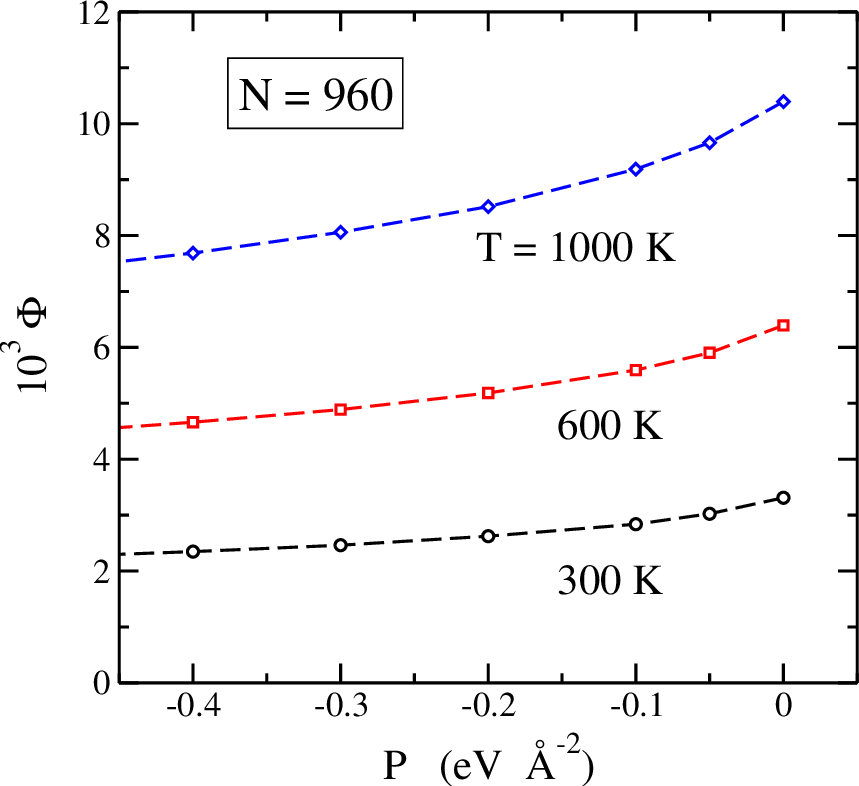}
\vspace{0.3cm}
\caption{Excess area per atom vs 2D hydrostatic pressure $P$ for
cell size $N$ = 960 at $T$ = 300 K (circles), 600 K (squares),
and 1000 K (diamonds). Open symbols are data points derived from
MD simulations of bilayer graphene. Dashed lines are guides to
the eye.
}
\label{f7}
\end{figure}

In Fig.~7 we display $\Phi$ for bilayer graphene as a function 
of 2D hydrostatic pressure $P$. Symbols represent data derived 
from our MD simulations at three temperatures: 
$T$ = 300 K (circles), 600 K (squares), and
1000~K (diamonds).  Dashed lines are guides to the eye.
These data were obtained for system size $N = 960$.
The excess area $\Phi$ increases as $T$ is raised,
in agreement with the growing amplitude of the out-of-plane
vibrational modes (see Eq.~ (\ref{omega3})).
In fact, a classical harmonic approximation (HA) for the vibrational 
modes predicts a linear increase of $\Phi$ with temperature.
From the results shown in Fig.~7 for $P = 0$, we find the
ratios $\Phi$(1000~K)/$\Phi$(300~K) = 3.14 and
$\Phi$(600~K)/$\Phi$(300~K) = 1.93, a little less than
the corresponding temperature ratios (3.33 and 2.0).
For a pressure $P = -0.5$ eV/\AA$^2$, we find
for those ratios at the same temperatures values of 3.28 and
1.99, respectively, closer to the harmonic expectancy. 
An in-plane tensile stress causes a decrease
in the vibrational amplitudes in the $z$ direction, and then the
modes are better described by a harmonic approach.

\section{Elastic constants and compressibility at finite temperatures}

\subsection{Temperature dependence}

Using MD simulations one can gain insight into the elastic
properties of materials under different kinds of applied
stresses, e.g., hydrostatic or uniaxial. In particular,
we consider elastic stiffness constants, $c_{ij}$, and 
compliance constants, $s_{ij}$, for 2D crystalline materials 
with hexagonal structure such as graphene.
We call $\tau_{ij}$ and $e_{ij}$ the components of the stress 
and strain tensors, respectively.
$\tau_{ij}$ is the force per unit length parallel to direction
$i$, acting in the $xy$ plane on a line  perpendicular to 
the $j$ direction.
We use the standard notation for strain components,
with $e_{ij} = \epsilon_{ij}$ for $i = j$, and
$e_{ij} = 2 \epsilon_{ij}$ for $i \neq j$ \cite{as76,ma18}.
More details on elastic properties of 2D crystals can be found in
Ref.~\cite{be96b}.

In terms of the compliance constants, we have for applied stress
$\{ \tau_{ij} \}$:
\begin{equation}
\begin{pmatrix}
  e_{xx} \cr  e_{yy}  \cr  e_{xy}
\end{pmatrix}
=
\begin{pmatrix}
s_{11} & s_{12} &   0     \cr
s_{12} & s_{11} &   0     \cr
  0    &   0    &  2 (s_{11}-s_{12})
\end{pmatrix}
\begin{pmatrix}
\tau_{xx}  \cr \tau_{yy}  \cr \tau_{xy} 
\end{pmatrix}
\; .
\label{sij}
\end{equation}
The matrix of stiffness constants $\{ c_{ij} \}$ is the inverse 
of $\{ s_{ij} \}$, so that we have the relations
\begin{eqnarray}
c_{11} & = & \frac{s_{11}} {s_{11}^2 - s_{12}^2}  \; ,  \label{c11b}  \\
c_{12} & = & \frac {s_{12}} {s_{12}^2 - s_{11}^2}  \; .  \label{c12b}
\end{eqnarray}

\begin{figure}
\includegraphics[width=7cm]{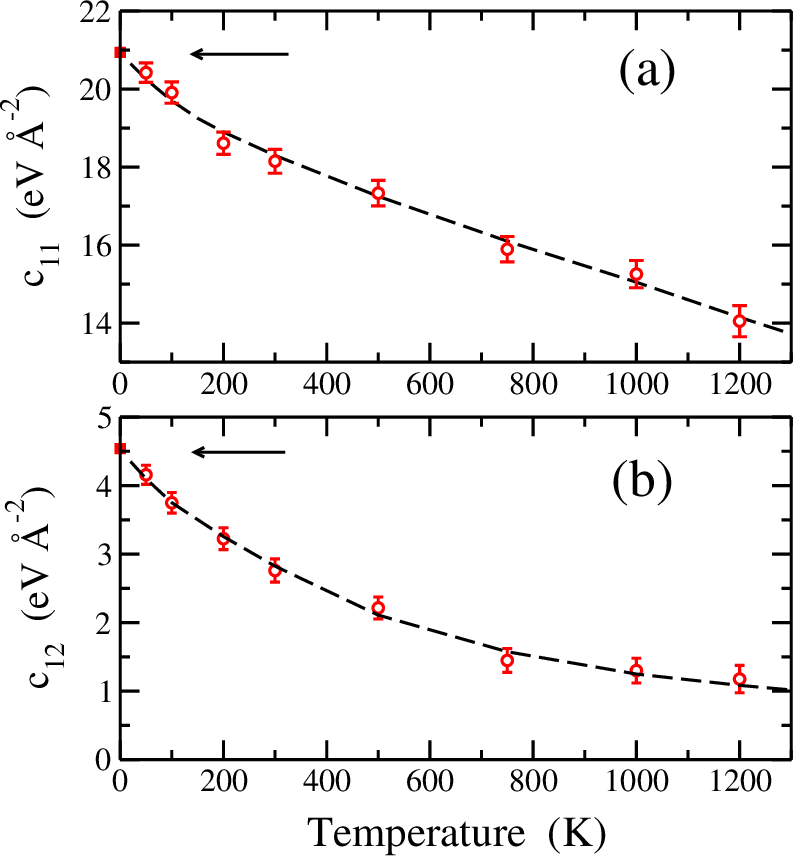}
\vspace{0.3cm}
\caption{Temperature dependence of the elastic stiffness constants
of bilayer graphene,
as derived from MD simulations for $N$ = 960 (open circles):
(a) $c_{11}$, (b) $c_{12}$. Dashed lines are guides to the eye.
Solid squares at $T = 0$, indicated by arrows, show the values of
$c_{11}$ and $c_{12}$ derived from the HA
using Eqs.~(\ref{c11}) and (\ref{c12}).
}
\label{f8}
\end{figure}

In Fig.~8 we present the stiffness constants as a function of
temperature, as derived from our MD simulations of bilayer graphene,
using Eq.~(\ref{sij}) (open circles). 
Panels (a) and (b) show results for $c_{11}$ and $c_{12}$, respectively.
Solid squares at $T$ = 0, signaled by arrows, indicate results for 
$c_{11}$ and $c_{12}$ obtained from the phonon dispersion bands as 
indicated in Sec.~II, using Eqs.~(\ref{c11}) and (\ref{c12}).
We find that finite-temperature data for the stiffness constants 
converge at low $T$ to the results of the HA
for both $c_{11}$ and $c_{12}$.
For rising temperature, the stiffness constants decrease rather fast.
This decrease is especially large for $c_{12}$, which is
found to be 1.18~eV/\AA$^2$ at $T$ = 1200~K vs the classical low-$T$ 
limit of 4.54~eV/\AA$^2$.

Comparing the elastic constants $c_{11}$ and $c_{12}$ found here
for bilayer graphene with those corresponding to monolayer 
graphene \cite{he17} and graphite \cite{he21b} (normalized to one layer), 
we find that they increase for the sequence monolayer-bilayer-graphite. 
This agrees with the fact that interaction between layers reduces 
the amplitude of out-of-plane vibrational modes, thus favoring 
an increase in the ``hardness'' of the layers. This trend is similar
to that discussed below for the 2D compression modulus $B_p$.

The Poisson's ratio $\nu$ can be obtained as the quotient
$c_{12} / c_{11}$. This yields for $T= 0$ (HA) $\nu = 0.22$. 
From the results of our simulations, we find
$\nu$ = 0.15 and 0.09 for $T$ = 300 and 1000~K, respectively,
with an important reduction for rising $T$, as a consequence of 
the decrease in $c_{12}$.
Calculations based on the self-consistent screening 
approximation \cite{do92,ko13,do18b} (SCSA) predict for $P = 0$ in the 
thermodynamic limit ($N \to \infty$) a Poisson's ratio $\nu = -1/3$. 
A negative value for this ratio is also expected from the
calculations presented by Burmistrov {\em et al.} \cite{bu18}
for $N \to \infty$. From the results of our MD simulations, 
we do not find, however, any indication for a negative Poisson's
ratio in the parameter region considered here. This is in line
with earlier results of Monte Carlo simulations by 
Los~{\em et al.} \cite{lo16} for monolayer graphene in a region
of system sizes larger than those considered here for 
bilayer graphene.

The 2D modulus of hydrostatic compression $B_p$ is defined 
for layered materials at temperature $T$ as \cite{be96b}
\begin{equation}
  B_p = - \frac{S_p}{n}  
	\left( \frac{\partial P}{\partial S_p} \right)_T   \, .
\label{bulk1}
\end{equation}
Note the factor $n$ (number of sheets) in the denominator, i.e.,
$B_p$ is the compression modulus per layer.
$P$ and $S_p$ appearing on the r.h.s. of Eq.~(\ref{bulk1})
are variables associated to the layer plane, and in fact the
pressure in the isothermal-isobaric ensemble used here is 
the conjugate variable to the area $S_p$.

One can also calculate the modulus $B_p$ on the basis of
the fluctuation formula \cite{la80,ra17,he18b}
\begin{equation}
       B_p = \frac{k_B T S_p}{n N (\Delta S_p)^2}   \; ,
\label{bulk2}
\end{equation}
where $(\Delta S_p)^2$ is the mean-square fluctuation of the area 
$S_p$, which is calculated here from MD simulations at $P = 0$.
This formula provides us with a practical procedure to obtain
$B_p$, vs calculating the derivative $(\partial S_p / \partial P)_T$
by numerical methods, which requires additional MD simulations at 
hydrostatic pressures close to $P = 0$.
For some temperatures we have verified that results for $B_p$ 
found with both procedures coincide within statistical error bars,
which is a consistency check for our results.

The modulus $B_p$ can be also obtained from the elastic constants
of bilayer graphene. Taking into account Eq.~(\ref{sij}),  
the change of the in-plane area, $\Delta S_p$ due to a 2D 
hydrostatic pressure $P$, is given by
\begin{equation}
 \frac{\Delta S_p}{S_p} = e_{xx} + e_{yy} = - 2 (s_{11} + s_{12}) P  \; .
\label{dap}
\end{equation}
Combining Eqs.~(\ref{bulk1}) and (\ref{dap}), one finds
\begin{equation}
	B_p = \frac {1}{2 (s_{11} + s_{12})}   \; ,
\label{bulk3}
\end{equation}
which can be also written as $B_p = (c_{11} + c_{12}) / 2$.
These expressions are valid for 2D materials with hexagonal symmetry.

\begin{figure}
\includegraphics[width=7cm]{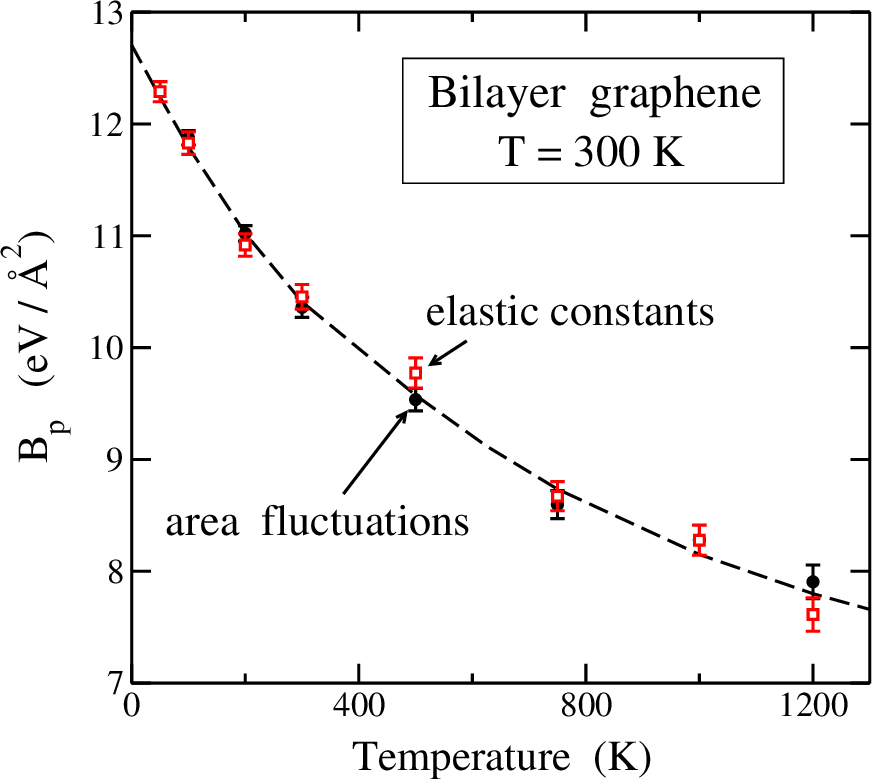}
\vspace{0.3cm}
\caption{2D modulus of hydrostatic compression, $B_p$, as a function of
temperature for stress-free bilayer
graphene ($P$ = 0) and $N$ = 960. Solid circles represent results
derived from fluctuations of the in-plane area, using Eq.~(\ref{bulk2}).
Open squares are data points obtained from the elastic constants
$s_{11}$ and $s_{12}$.
The dashed line is a guide to the eye.
}
\label{f9}
\end{figure}

In Fig.~9 we present the modulus $B_p$ of bilayer graphene
as a function of $T$, as derived from our MD simulations.
Solid circles represent results obtained from the in-plane area 
fluctuation $(\Delta S_p)^2$ by employing Eq.~(\ref{bulk2}).
Open squares are data points calculated from the elastic
constants. Both sets of results coincide within error bars.
At low temperature, $B_p$ converges to the value given by 
the expression
\begin{equation}
  B_0 =  \frac{S_p}{n}  \frac{\partial^2 E}{\partial S_p^2}  \; ,
\label{bulk0}
\end{equation}
where $E$ is the energy. For bilayer graphene we have
$B_0$ = 12.74 eV/\AA$^2$, which agrees with the extrapolation
of finite-$T$ results to $T = 0$. 
The modulus $B_p$ derived from MD simulations is found to decrease 
fast as the temperature is raised, and at $T$ = 1200 K it amounts 
to about 60\% of the low-$T$ limit $B_0$.

For monolayer graphene, the same interatomic potential
yields in a HA: $B_0$ = 12.65 eV/\AA$^2$, somewhat less than 
the value found for the bilayer.
This difference is larger at finite temperatures.
Even though interlayer interactions are relatively weak, they give 
rise to a reduction in the vibrational amplitudes of out-of-plane 
modes, and as a result the graphene sheets become ``harder'' 
in the bilayer, so that the modulus $B_p$ increases with respect 
to an isolated sheet.  Moreover, the difference
between the modulus $B_p$ per sheet for bilayer and monolayer
graphene grows for rising system size $N$ (see Sec.~VII).

The in-plane Young's modulus $E_p$ can be obtained from $B_p$ through 
the expression $E_p = 2 B_p (1 - \nu)$. This yields for the bilayer
at $T = 0$, $E_p$ = 19.87 eV/\AA$^2$, which translated into units of
force per square length gives $E_p / d =$ 0.954 TPa, similar 
to values appearing in the literature \cite{me15,ca18c}.

\subsection{Mechanical instability under stress}

\begin{figure}
\includegraphics[width=7cm]{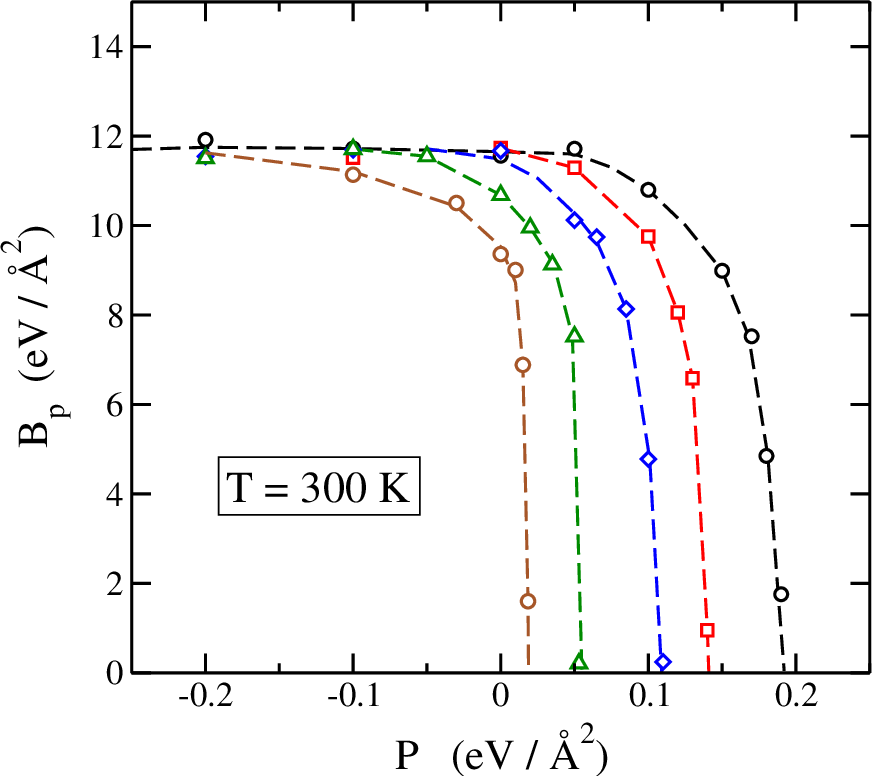}
\vspace{0.3cm}
\caption{Modulus $B_p$ as a function of 2D hydrostatic pressure
at $T$ = 300 K, for various cell sizes. Symbols are data points
derived from MD simulations for various cell sizes.
From left to right: $N$ = 8400, 960, 448, 308, and 240.
Dashed lines are guides to the eye.
}
\label{f10}
\end{figure}

The modulus $B_p$ is particularly interesting to study the critical
behavior of bilayer graphene under 2D hydrostatic pressure.
In Fig.~10 we present the dependence of $B_p$ on $P$, including
tensile and compressive stresses. Symbols represent values derived
from MD simulations for various cell sizes. 
From left to right: $N$ = 8400, 960, 448, 308, and 240.
For each size $N$, increasing in-plane compression causes a fast 
decrease in $B_p$, which vanishes for a pressure $P_c(N)$, where 
the bilayer graphene with planar sheets becomes mechanically unstable. 
This is typical of a spinodal point in the $(P, T)$ phase 
diagram \cite{sc95,he03b,ra18b,ca85}.
For $P > P_c$, the stable configuration corresponds to wrinkled
graphene sheets, as observed earlier for monolayer graphene \cite{ra17}.

For given $N$ and $T$, there is a pressure region (compressive stress)
where bilayer graphene is metastable, i.e., for $P < P_c$. 
The spinodal line, which delineates the metastable phase from the unstable
phase, is the locus of points $P_c(N,T)$ where $B_p = 0$.
This kind of spinodal lines have been studied earlier
for water \cite{sp82}, as well as for ice, SiO$_2$ cristobalite \cite{sc95},
and noble-gas solids \cite{he03b} near their stability limits.
In recent years, this question has been investigated for 2D materials,
and in particular for monolayer graphene \cite{ra18b,ra20}. 

According to Eq.~(\ref{bulk2}), vanishing of $B_p$ 
for finite $N$ corresponds to a divergence of the area fluctuation
$(\Delta S_p)^2$ to infinity. Moreover, the MSF of the atomic
$z$ coordinate, $(\Delta h)^2$, diverges at the corresponding spinodal
pressure, as mentioned in Sec.~IV.C.   For graphene bilayers,
this spinodal instability is associated to a soft 
vibrational mode in the ZA branch. In fact, for each $N$ this 
instability appears for increasing $P$ when the frequency of 
the ZA vibrational mode with minimum wavenumber, $k_{\rm min}$, 
reaches zero ($\omega_{\rm min} \to 0$).

Close to a spinodal point, the modulus $B_p$ behaves as a function
of $P$ as $B_p \sim (P_c - P)^{1/2}$ (see Appendix C).
This pressure dependence agrees with the shape of the curves shown 
in Fig.~10 near the spinodal pressure $P_c$ for each size $N$.  
Note that $P_c$ moves to smaller compressive pressures as $N$ rises. 
This size effect is analyzed below in Sec.~VII.
We also note that, for a given size $N$, the critical stress $P_c$ 
depends on temperature, as shown before for monolayer 
graphene \cite{ra18}. It was found that $P_c$ increases for rising
temperature, as a consequence of a raise in vibrational amplitudes
in the out-of-plane direction. We have checked that something similar 
happens for bilayer graphene, but a detailed study of this question 
requires additional MD simulations, which will be the subject of 
future work.

\section{Size effects}

\begin{figure}
\includegraphics[width=7cm]{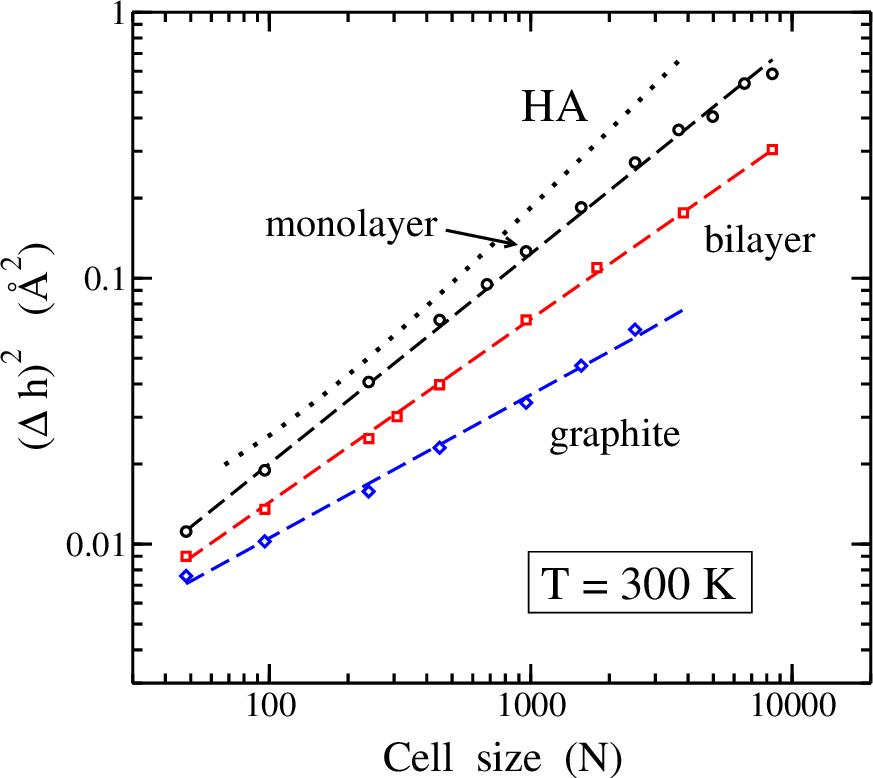}
\vspace{0.3cm}
\caption{Atomic MSF in the $z$ direction vs cell size for $T$ = 300 K
in a logarithmic plot.
Data derived from MD simulations are given for
monolayer graphene (circles), bilayer graphene (squares), and
graphite (diamonds). Error bars are in the order of the symbol size.
Dashed lines are least-square fits to the data points.
The dotted line represents the MSF corresponding to a HA.
}
\label{f11}
\end{figure}

As noted above, some properties of 2D materials display 
important size effects. In this section, we concentrate on the
size dependence of the MSF $(\Delta h)^2$, the modulus $B_p$,
and the spinodal pressure $P_c$ for bilayer graphene, and 
study their asymptotic behavior for large $N$.

In Fig.~11 we present in a logarithmic plot the atomic MSF 
$(\Delta h)^2$ in the $z$ direction as a function of system size 
at $T$ = 300~K.  Results derived from MD simulations for 
stress-free bilayer graphene are shown as open squares.
For comparison we also display data for monolayer graphene (circles)
and graphite (diamonds), obtained from MD simulations with the
LCBOPII interatomic potential.
For the system sizes presented in Fig.~11, $(\Delta h)^2$ may be 
expressed in the three cases as a power of $N$: 
$(\Delta h)^2 \sim N^{\alpha}$.
We find for the exponent $\alpha$ values of 0.78, 0.69, and 0.56
for monolayer, bilayer graphene, and graphite, respectively.

The MSF in the $z$ direction can be written in a HA as:
\begin{equation}
    (\Delta h)^2  =   \frac{k_B T}{N \rho S_p}
        \sum_{j,{\bf k}} \frac{1}{\omega_j({\bf k})^2} \; ,
\label{deltah2}
\end{equation}
where the sum in $j$ is extended to the phonon bands with atomic
motion in the out-of-plane direction, i.e., ZA, ZO, and ZO' for
bilayer graphene (as above in Eq.~(\ref{omega3})).
The sum in Eq.~(\ref{deltah2}) is dominated by ZA modes with
wavevector close to the $\Gamma$ point, i.e. small frequency 
$\omega$.  The inputs of bands ZO and ZO' are almost independent 
of the system size, and they give a joint contribution of
$\approx 8 \times 10^{-3}$~\AA$^2$ to $(\Delta h)^2$ in 
Eq.~(\ref{deltah2}).

For the ZA band, putting a dispersion $\rho \, \omega^2 = \kappa k^4$,
one finds \cite{ga14}
\begin{equation}
 (\Delta h)^2_{\rm ZA} = \frac{C \, k_B T S_p N}{16 \pi^4 \, \kappa}
	\; ,
\end{equation}
where $C = 6.03$ is a constant. Thus, in a HA the ZA band 
yields a contribution proportional to $N$ 
and an exponent $\alpha = 1$.
The result of the HA including the inputs of the three phonon
bands ZA, ZO, and ZO' is shown in Fig.~11 as a dotted line. 
For large $N$, $(\Delta h)^2$ is dominated by atomic displacements
associated to ZA modes and we find that it increases linearly 
with $N$.  For small $N$, one observes a departure
from the linear trend due to contributions of ZO and ZO' modes.

The size dependence of $(\Delta h)^2$ obtained from MD simulations
for stress-free bilayer graphene can be understood assuming an
effective dependence for the frequency of ZA modes as
$\rho \, \omega^2 = \bar{\kappa} \, k^{\beta}$, where
$\bar{\kappa}$ is a modified bending constant and
$\beta$ is an exponent controlling the frequency of 
long-wavelength (small frequency) modes.
This expression assumes an effective shape for the ZA band
at finite temperatures, as a consequence of anharmonicity
in the vibrational modes, and is similar to that considered
earlier for monolayer graphene \cite{ga14,tr13b,lo16}.
Assuming such a dispersion for ZA modes in the bilayer, we have
for large $N$:
\begin{equation}
 (\Delta h)^2_{\rm eff}  \approx  \frac{k_B T}{\bar{\kappa} N S_p}
     \sum_{{\bf k}} \frac{1}{k^{\beta}} \; ,
\label{deltah3}
\end{equation}
where the sum is extended to ${\bf k}$ points in the
2D Brillouin zone.

\begin{figure}
\includegraphics[width=7cm]{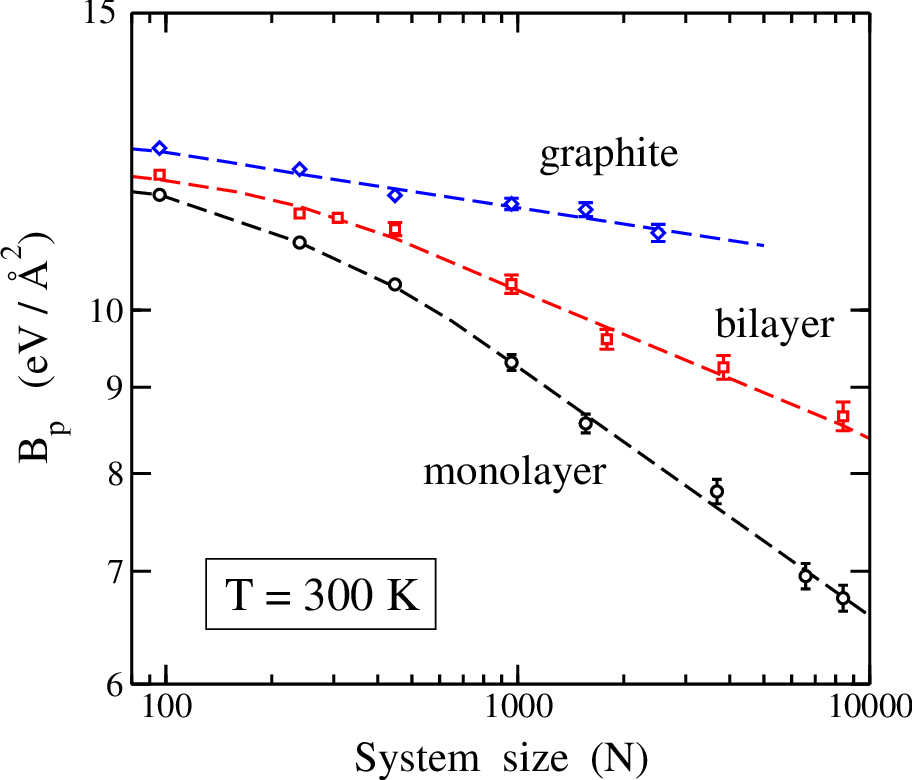}
\vspace{0.3cm}
\caption{Modulus $B_p$ as a function of cell size $N$
at $T$ = 300 K on a logarithmic plot.
Symbols represent data derived from MD simulations for
a graphene monolayer (circles), bilayer (squares), and
graphite (diamonds). Dashed lines are guides to the eye.
}
\label{f12}
\end{figure}

Taking into account the relation between the minimum
wavenumber $k_{\rm min}$ and the size $N$, and replacing
the sum in Eq.~(\ref{deltah3}) by an integral, one finds
a size dependence
\begin{equation}
   (\Delta h)^2_{\rm eff}  =  \frac{D}{\bar{\kappa}}  
	k_B T  \, (S_p  N)^{\frac{\beta}{2} - 1} \; .
\label{deltah4}
\end{equation}
where $D$ is an integration constant.
This means that our exponent $\alpha$ can be related to 
$\beta$ as $\alpha = \beta / 2 - 1$, which yields
for bilayer graphene $\beta$ = 3.38. 
Similar effective exponents can be derived from MD simulation
results for monolayer graphene and graphite, for which
we find $\beta$ = 3.56 and 3.12, respectively.

The 2D modulus of hydrostatic compression $B_p$ introduced in
Sec.~VI also displays finite-size effects.
In Fig.~12 we show in a logarithmic plot the dependence of $B_p$
on $N$ at $T$ = 300~K.
Open squares represent results obtained for bilayer graphene
from MD simulations. For comparison, we also display data
for monolayer graphene (circles), as well as for graphite 
(diamonds).
For $N > 500$, $B_p$ can be fitted for the bilayer to an expression
$B_p \sim N^{-\zeta}$, with an exponent $\zeta = 0.086$.
From similar fits for monolayer graphene and graphite, we find
the exponents 0.159 and 0.033, respectively.
Note that the exponent $\zeta$ for the bilayer is about one
half of that corresponding to the monolayer.
This indicates that the size effect is less important
for the former than for the latter, as visualized in Fig.~12.

Looking at Eq.~(\ref{bulk2}), and taking into account that
$S_p$ changes slowly with $N$, we have for the area fluctuation 
a size dependence: $(\Delta S_p)^2 \sim N^{\zeta - 1}$.
This means that $(\Delta S_p)^2_B / (\Delta S_p)^2_M \sim N^{-0.073}$,
where the subscripts $B$ and $M$ refer to bilayer and monolayer,
respectively. For small $N$, the area fluctuation is similar for 
bilayer and monolayer graphene, but they become comparatively smaller
for the bilayer as the size increases.
Our exponent for the monolayer can be translated to a dependence
on the cell side length $B_p \sim L^{- 2 \zeta}$, with 
$2 \zeta$ = 0.318, close to the exponent 0.323 found by
Los {\em et al.} \cite{lo16} for the dependence of $B_p$ on $L$. 

The size dependence of the critical pressure $P_c$ introduced 
in Sec.~VI is shown in Fig.~13, where we have plotted values of 
$P_c$ derived from MD simulations at $T$ = 300~K for several
cell sizes. One observes at first sight a linear dependence of
$P_c$ with the inverse cell size, $N^{-1}$.
This dependence may be understood by considering the effect of
a compressive stress on ZA vibrational modes, as follows.

For a single graphene layer under a 2D hydrostatic pressure $P$,
the dispersion relation of ZA modes may be written as
\begin{equation}
  \rho \, \omega^2 = \sigma k^2 + \kappa \, k^4
\label{room2}
\end{equation}
where $\sigma = -P$ \cite{he17,ra18b}. For increasing compressive 
stress, $\omega$ is reduced, i.e. $d \omega / d P < 0$.
Thus, for system size $N$,
a graphene layer becomes unstable when the frequency of the
ZA mode with wavenumber $k_{\rm min}$ vanishes. This occurs for
\begin{equation}
    \sigma \, k_{\rm min}^2 + \kappa \, k_{\rm min}^4 = 0   \; ,
\end{equation}
which yields a critical stress 
\begin{equation}
     \sigma_c =  - \kappa \, k_{\rm min}^2  \; .
\end{equation}
As indicated above, the minimum wavenumber $k$ present for cell
size $N$ is $k_{\rm min} = 2 \pi / (N S_p)^{1/2}$.
For bilayer graphene, the in-plane critical pressure is 
given by $P_c = - 2 \sigma_c$, from where we have:
\begin{equation}
     P_c \, N =  \frac {8 \, \pi^2 \kappa}{S_p}   \; .
\label{pcn}
\end{equation}
Putting $\kappa$ = 1.49 eV and $S_p$ = 2.617 \AA$^2$/atom,
we find $P_c \, N$ = 44.95 eV/\AA$^2$, which is the line displayed 
in Fig.~13. 
This line matches well the results of our MD simulations
(solid circles), with the exception of the result for the largest
size presented in the figure. This deviation from the general trend
of smaller sizes may be due to three reasons. 
First, the presence in the graphene layers of a residual (small)
intrinsic stress at $T = 300$~K, which is not detected for small
$N$, due to the larger values of the corresponding pressure
$P_c$ \cite{ra16,ra18}.
Second, the graphene bilayer can remain in a metastable state 
along millions of MD simulation steps for large cell size. 
This means that observation of the true transition (spinodal) point
could require much longer simulations, not available at present
for such large system sizes.
Third, the dispersion relation for the ZA band in Eq.~(\ref{room2}),
utilized to obtain Eq.~(\ref{pcn}), may be modified for small
$k$ (long wavelength), so that the exponent 4 on the r.h.s.
could be renormalized in a similar way to the exponent $\beta$
found for the size dependence of $(\Delta h)^2$.

\begin{figure}
\includegraphics[width=7cm]{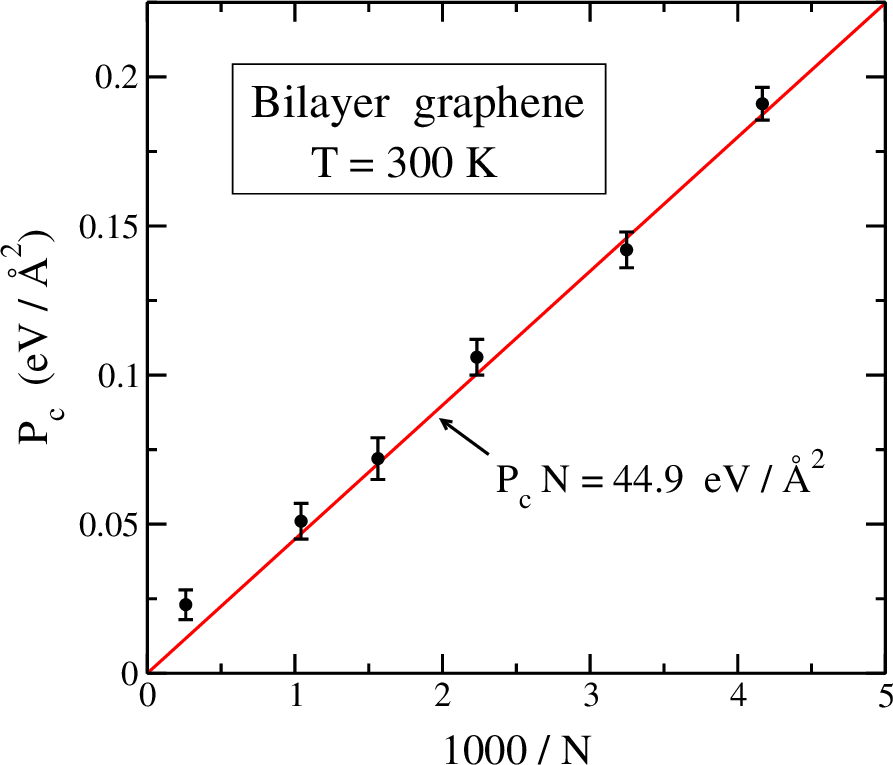}
\vspace{0.3cm}
\caption{Critical pressure $P_c$ as a function of the inverse
cell size at $T$ = 300 K. Solid circles represent data derived
from MD simulations of bilayer graphene under in-plane stress.
The solid line was calculated using Eq.~(\ref{pcn}).
}
\label{f13}
\end{figure}

Calculations based on the SCSA predict a universal behavior for
scaling exponents in the thermodynamic limit 
($N \to \infty$) \cite{do92,ko13,do18b}.
This means that the exponents presented above should
coincide for 2D systems (including monolayers and bilayers)
in the large-size limit. 
According to such calculations, universality is approached for
system size larger than a crossover scale given by the so-called
Ginzburg length, $L_G$. This temperature-dependent length can be
estimated for graphene (using the bending constant $\kappa$ and 
Young's modulus $E_p$) to be around $40-50$~\AA\ at 
$T = 300$~K \cite{an12,ma20}. This corresponds in our notation
to a system size $N_G \sim 900$.
For bilayer graphene, we have considered here simulation cells 
with length sides up to 150~\AA, well above those values of $L_G$.
One can, however, understand $L_G$ as a reference length for 
the crossover to a regime where universality is approached, 
and a direct detection of this universality could be only found 
for lengths clearly larger than $L_G$.
In any case, from the results of our simulations for bilayer 
(with $L$ up to 150~\AA) and monolayer graphene presented here, 
we do not find any evidence or trend indicating that such a kind 
of universality will appear for results derived from simulations 
using larger cells. Thus, if such kind of universality is in fact
a physical aspect of 2D crystalline membranes, as predicted by 
the SCSA, it has not yet been observed from atomistic simulations 
with interaction potentials mimicking those of actual materials,
as graphene.

\section{Summary}

MD simulations allow us to gain insight into elastic properties of
2D materials, as well as on their stability under external stress.
We have presented here the results of extensive simulations of
bilayer graphene using a well-checked interatomic potential, for
a wide range of temperatures, in-plane stresses, and system sizes.
We have concentrated on physical properties such as the excess area,
interlayer spacing, interatomic distance, elastic constants, 
in-plane compression modulus, and atomic MSF 
in the out-of-plane direction.

The elastic constants are found to appreciably change as
a function of temperature, especially $c_{12}$. This
causes a reduction of the Poisson's ratio for rising $T$.
The in-plane compression modulus $B_p$ has been obtained from 
the fluctuations of the in-plane area, a procedure which
yields results consistent with those derived from the
elastic constants of the bilayer.

For bilayer graphene under in-plane stress, we find a divergence 
of the MSF $(\Delta h)^2$ for an in-plane pressure $P_c(N)$,
which corresponds to the limit of mechanical stability of
the material. This divergence is accompanied by
a vanishing of the in-plane compression modulus,
or a divergence of the compressibility $\xi_p = 1/ B_p$.

Finite-size effects are found to be important for several
properties of bilayer graphene.  The spinodal pressure $P_c$
is found to scale with system size as $1/N$.
A similar scaling with the inverse size is obtained for
the MSF of the interlayer spacing: $(\Delta d)^2 \sim N^{-1}$.
The atomic out-of-plane MSF also follows 
a power law  $(\Delta h)^2 \sim N^{\alpha}$  with an
exponent $\alpha$ = 0.69.
For $B_p$, we find for $N > 500$ at $T = 300$~K a dependence
$B_p \sim N^{-\zeta}$, with an exponent $\zeta = 0.086$.

Comparing the simulation results with those obtained from
a HA gives insight into finite-temperature anharmonic effects.
Thus, for the atomic MSF in the
out-of-plane direction, a HA predicts a linear dependence
of $(\Delta h)^2$ with system size $N$, to be compared with
the sublinear dependence obtained from the simulations.

The change with system size $N$ of $(\Delta h)^2$ and 
the modulus $B_p$ for bilayer graphene is slower 
(i.e., less slope in Figs. 11 and 12) than for the monolayer.
This is indeed due to interlayer interactions, which
manifests themselves in the presence of the layer-breathing 
ZO' phonon branch. According to calculations based on the
SCSA, the size dependence of physical observables such as
$(\Delta h)^2$ or the in-plane modulus $B_p$ should be
controlled, for large $N$, by universal exponents independent 
of the particular details of the considered 2D system (monolayer 
or bilayer graphene in our case). 
We have not observed this universality from our
MD simulations for cell size up to 150~\AA, and a clarification
of this question remains as a challenge for future research.

We finally note that MD simulations as those presented here 
can give information on the properties of graphene multilayers 
under stress.  This may yield insight into the relative 
stability of such multilayers in a pressure-temperature 
phase diagram. Moreover, nuclear quantum effects can affect
the mechanical properties of graphene bilayers and multilayers
at low temperatures, as shown earlier for graphite. This
question can be addressed using atomistic simulations with
techniques such as path-integral molecular dynamics.  \\ \\

\begin{acknowledgments}
J. H. Los is thanked for his help in the implementation of the 
LCBOPII potential and for providing the authors with an updated
version of its parameters.
The research leading to these results received funding from 
Ministerio de Ciencia e Innovaci\'on (Spain) under Grant Numbers
PGC2018-096955-B-C44 and PID2022-139776NB-C66.
\end{acknowledgments}


\appendix

\section{Updates of the LCBOPII model}

\subsection{Torsion term}

The torsion term $t_{ij}$ in the original LCBOPII model \cite{lo05}
was modified to improve the value of the bending rigidity, $\kappa$, 
in graphene. Eq. (36) in the original paper was changed in 
the following way:
\begin{equation}
  t_{ij}(\tilde{y},\tilde{z}) = 
   (1 - \tilde{z}^3) (1 + \tilde{z} (1 + \tilde{z}) \chi(\tilde{y})) 
	\tau_0(\tilde{y}) + \tilde{z}^3 \tau_1 (\tilde{y})  \: ,
\end{equation}
where the variables $\tilde{y},\tilde{z}$ are defined as in 
Ref.~\cite{lo05}.   The new functions are:
\begin{equation}
  \tau_0(\tilde{y}) = A_{t1} + A_{t2} \tilde{y}^2 (3 - 2\tilde{y}^2)  \: ,
\end{equation}
\begin{equation}
  \tau_1(\tilde{y}) = \frac{B_{t1} + B_{t2} \tilde{y}^2 + 
	B_{t3} \tilde{y}^4} {1 + B_{t4}\tilde{y}^2}  \: ,
\end{equation}
\begin{equation}
  \chi(\tilde{y}) = C_{t1} + C_{t2} \tilde{y}^2  \: .
\end{equation}
Values of the constants are: $A_{t1} = -0.049$, $A_{t2} = -0.022$, 
$B_{t1} = -0.295$, $B_{t2} = -3.361$, $B_{t3} = 1.150$, $B_{t4} = 19.616$,
$C_{t1} = 3.35$, $C_{t2} = -2.6$.
Units of energy and length are eV and \AA, respectively, as in
the original paper.

\subsection{Long-range-interaction}

The longe-range term $V^{lr}(r)$ of the original LCBOPII model \cite{lo05}
has been modified to improve the value of the interaction energy between 
both graphene layers in the bilayer.
The long-range interaction $V^{lr}(r)$
between two carbon atoms, given in Eq.~(42) of the original paper \cite{lo05},
is changed by a new one, but only when the two carbon atoms at a distance
$r$ are located at different graphene layers. The modified long-range
interaction has the following form:
\begin{eqnarray}
  V_{mod}^{lr}(r) = 
	\left(c_{1}e^{-\alpha(r-r_{0})} - \frac{c_{2}}{r^{6}} + 
       c_3 \frac{e^{-2\alpha(r-r_{0})}}{r^{6}}\right) S_{lr}^{down}(r) 
      \nonumber  \\  
        \text{if} \; r \leq 5.1 \; \text{\AA}   \nonumber
\end{eqnarray}
\begin{eqnarray}
  V_{mod}^{lr}(r) = \left[ (d_1+d_2r)(r-6)^2 \right] S_{lr}^{down}(r) \: ,
     \nonumber  \\   \text{if} \; 5.1 \; \text{\AA} \; < r < 6 \; 
	\text{\AA}  \: ,  \nonumber
\end{eqnarray}
\begin{equation}
  V_{mod}^{lr}(r) = 0, \hspace{0.5cm} \text{if} \;  r \geq 6 \; \text{\AA}  
	\: .    \nonumber
\end{equation}
The function $S_{lr}^{down}(r)$ smoothly cuts off the long-range
interactions at 6~\AA\ and it is defined in Tab.~I of 
Ref.~\cite{lo05}.  Values of the constants
are: $r_0 = 3.7157,\: c_1 = 3.0748 \times 10^{-3}, \: 
c_2 = 353.1877, \: c_3 = 334.9434, \; \alpha = 5.4767 \times 10^{-2} \: 
d_1 = -4.2249 \times 10^{-3}, \; \mathrm{and} \; d_2 = 6.1914 \times 10^{-4}$.
Units of energy and length are eV and \AA, respectively, as in
the original paper.

\section{Excess area}

A relation between the in-plane and real areas can be obtained
from a continuum description of a graphene sheet,
which considers the vibrational modes in the $z$ direction.  
The instantaneous real area per atom, $S_r^*$,
can be expressed in the continuum limit as \cite{im06,wa09,ra17}
\begin{equation}
   S_r^* = \frac1N \int_0^{L_x} \int_0^{L_y} dx \, dy \, 
	     \sqrt{1+|\nabla h({\bf r})|^2}  \; ,
\label{aap2}
\end{equation}
where ${\bf r} \equiv (x, y)$ is the 2D position and $h({\bf r})$ 
indicates the height of the surface, i.e. the distance to
the mean $xy$ plane of the sheet.
For small $|\nabla h({\bf r})|$, i.e., 
$(\partial h / \partial x)^2 + (\partial h / \partial y)^2 \ll 1$
(which is verified here), one has
\begin{equation}
  S_r^* \approx \frac1N \int_0^{L_x} \int_0^{L_y} dx \, dy \, 
	\left[ 1 + \frac12 |\nabla h({\bf r})|^2  \right]    \; .
\label{aap3}
\end{equation}

The out-of-plane displacement $h({\bf r})$ can be written as 
a Fourier series
\begin{equation}
    h({\bf r}) = \frac{1}{\sqrt{N}} 
        \sum_{\bf k} {\rm e}^{i {\bf k \cdot r}} H({\bf k})  \; ,
\label{hxy}
\end{equation}
with wavevectors ${\bf k} = (k_x, k_y)$ in the 2D hexagonal 
Brillouin zone, i.e., $k_x = 2 \pi n_x/ L_x$ and
$k_y = 2 \pi n_y/ L_y$ with integers $n_x$ and $n_y$ \cite{ra16}.
The Fourier components are given by
\begin{equation}
    H({\bf k}) = \frac{\sqrt{N}}{L_x L_y}  
       \int_0^{L_x} \int_0^{L_y} dx \, dy \,
       {\rm e}^{- i {\bf k \cdot r}}  h({\bf r})   \; .
\label{hk}
\end{equation}
The thermal average of the atomic MSF in the $z$-direction is 
given from the Fourier components by
\begin{equation}
  \langle h({\bf r})^2 \rangle =
	 \frac1N  \sum_{\bf k} \langle |H({\bf k})|^2 \rangle \; .
\label{hr2}
\end{equation}
From Eq.~(\ref{hxy}), we have
\begin{equation}
  \nabla h({\bf r})  =  \frac{i}{\sqrt{N}}
  \sum_{\bf k} {\bf k} \, {\rm e}^{i {\bf k \cdot r}}  H({\bf k}) \; ,
\label{nablah}
\end{equation}
and
\begin{equation}
  \langle |\nabla h({\bf r})|^2 \rangle  = 
    \frac1N  \sum_{\bf k}  k^2  \langle |H({\bf k})|^2 \rangle  \; ,
\label{nablah3}
\end{equation}
because $\langle H({\bf k}_1) H({\bf k}_2)^* \rangle = 0$ for
${\bf k}_1 \neq {\bf k}_2$.

Then, using Eqs.~(\ref{aap3}) and (\ref{nablah3}), the mean real 
area per atom can be written as
\begin{equation}
     S_r = \langle S_r^* \rangle  =  
	 S_p  \left( 1 + \frac{1}{2N}  \sum_{\bf k}  k^2  
	\langle |H({\bf k})|^2 \rangle \right)  \; ,
\end{equation}
with $S_p = L_x L_y / N$. For uncoupled vibrational modes in 
the out-of-plane $z$ direction (i.e., in a HA), 
$\langle |H({\bf k})|^2 \rangle$ may be expressed as a sum of MSFs:
\begin{equation}
  \langle |H({\bf k})|^2 \rangle = 
         \sum_j  \langle |\xi_j({\bf k})|^2 \rangle
\label{hk2}
\end{equation}
with 
$\langle |\xi_j({\bf k})|^2 \rangle = k_B T / m \omega_j({\bf k})^2$,
so that we have for the excess area:
\begin{equation}
 \Phi  = \frac{S_r - S_p}{S_p} =   \frac{k_B T}{2 m N}  
	\sum_{j,{\bf k}} \frac{k^2}{\omega_j({\bf k})^2} \; .
\label{omega2}
\end{equation}
The sum in $j$ in Eqs.~(\ref{hk2}) and (\ref{omega2}) runs over 
the phonon bands with atom displacements in the $z$ direction
(ZA, ZO', and the two-fold degenerate ZO branch).
We note that the in-plane area fluctuates in our simulations, but
its fluctuations are not considered in the harmonic calculation
presented here.

\section{Spinodal pressure}

Close to a spinodal point, the free energy at temperature $T$ 
can be written as \cite{ma91,bo94b,ra20}
\begin{equation}
   F = F_c + a_1 \, (S_p  - S_p^c) + a_3 \, (S_p  - S_p^c)^3 + ...  \; ,
\label{ffc}
\end{equation}
where $F_c$ and $S_p^c$ are the free energy and in-plane area
at the spinodal point. At this point one has
$\partial^2 F / \partial S_p^2 = 0$, so that a quadratic term
does not appear on the r.h.s of Eq.~(\ref{ffc}), i.e., $a_2 = 0$.
The coefficients $a_i$ are in general dependent of the temperature. 

The pressure is
\begin{equation}
   P = - \frac{\partial F}{\partial S_p} =
          - a_1 - 3 a_3 \, (S_p  - S_p^c)^2 + ...  \; ,
\label{pfap}
\end{equation}
and $P_c = - a_1$ is the spinodal pressure, as it corresponds
to the area $S_p^c$.
The 2D modulus of hydrostatic compression is given by
\begin{equation}
  B_p = \frac{S_p}{n} \, \frac{\partial^2 F}{\partial S_p^2} =
     - \frac{S_p}{n}  \, \frac{\partial P}{\partial S_p}   \; .
\label{bpa}
\end{equation}
where $n$ is the number of sheets in the 2D material.
Then, to leading order in an expansion in powers of $S_p - S_p^c$,
we have
\begin{equation}
  B_p = \frac{6 a_3}{n} S_p^c (S_p - S_p^c)   \; ,
\end{equation}
or, considering Eq.~(\ref{pfap}),
\begin{equation}
  B_p = \frac{2}{n} \sqrt{3 a_3} \, S_p^c \, (P_c - P)^{1/2} \; .
\label{bpa2}
\end{equation}



\end{document}